\documentclass[11pt,a4paper]{article}
\pdfoutput=1

\usepackage[english]{babel}
\usepackage[left=0.8in,right=0.8in,top=1.0in,bottom=1.2in]{geometry}

\usepackage{dsfont}
\usepackage{amsfonts}
\usepackage{amsmath}
\usepackage{amssymb}
\usepackage{amsthm}
\usepackage{bbold}
\usepackage{calc}
\usepackage{cancel}
\usepackage{float}
\usepackage{slashed}
\usepackage{mathrsfs}
\usepackage{pdfpages}
\usepackage{hyperref}
\usepackage{multirow}
\usepackage{mathtools}
\usepackage{cite}
\usepackage{tikz}
\usepackage{tikzsymbols}

\usepackage{array}

\usepackage{colortbl} % Always load before 'arydshln'. Otherwise, vertical dashed lines in tables disappear
\definecolor{Gray}{gray}{0.95}
\definecolor{RGray}{gray}{0.90}
\definecolor{CGray}{gray}{0.92}

\usepackage{arydshln}

\numberwithin{equation}{section}
\numberwithin{figure}{section}
\numberwithin{table}{section}

\allowdisplaybreaks

\newcommand{\gsim}{\lower.7ex\hbox{$\;\stackrel{\textstyle>}{\sim}\;$}}
\newcommand{\lsim}{\lower.7ex\hbox{$\;\stackrel{\textstyle<}{\sim}\;$}}

\makeatletter
\g@addto@macro\bfseries{\boldmath}
\makeatother

\begin{document}
%%%%%%%%%%%%%%%%%%%%%%%%%%%%%%%%%%%%%%%%%%%%%%%%%%%%%%%%%%%%%%%%%%

\begin{flushright}
 ZU-TH-15/19  \\
\end{flushright}

\begin{center}
\vspace{0.7cm}
     {\Large\bf Revisiting the vector leptoquark explanation\\[0.2 cm] of the $B$-physics anomalies}
       \\ [1.0 cm] 
   {\bf Claudia Cornella,  Javier Fuentes-Mart\'{\i}n, Gino Isidori}    \\[0.2cm]
  {\em Physik-Institut, Universit\"at Z\"urich, CH-8057 Z\"urich, Switzerland}  \\
\end{center}
\vspace{0.5 cm}

\centerline{\large\bf Abstract}
\begin{quote}
We present a thorough investigation of the vector leptoquark hypothesis for 
a combined explanation of  the $B$-physics anomalies. We analyze this hypothesis 
from a twofold perspective, taking into account recent results from $B$-physics observables 
and high-$p_T$ searches. 
 First, using a simplified model, 
we determine the general conditions for a  successful low-energy fit 
in presence of right-handed  leptoquark couplings (neglected in previous analyses).
Second, we show how  these conditions, in particular a sizable 2-3 family mixing, 
can be achieved in a  motivated ultraviolet completion.
Our analysis reinforces the phenomenological success of the vector leptoquark hypothesis 
in addressing the anomalies, and its compatibility with
motivated extensions of the Standard Model based on the idea of flavor non-universal gauge interactions.
The implications of right-handed leptoquark couplings for a series of key low-energy observables, 
namely $B_s \to \tau\tau$ and $\tau\to\mu$ 
lepton flavor violating processes, both in $\tau$ and in $B$ decays, 
are discussed in detail. The role of the  ultraviolet completion in precisely 
estimating other low-energy observables, most notably $\Delta F=2$ amplitudes,
is also addressed.
\end{quote}

\thispagestyle{empty}
\setcounter{page}{0}

\tableofcontents
\newpage 

%%%%%%%%%%%%%%%%%%%%%%%%%%%%%%%%%%%%%%%%%%%%%%%%%%%%%%%%%%%%%%%%%%
\section{Introduction}
%%%%%%%%%%%%%%%%%%%%%%%%%%%%%%%%%%%%%%%%%%%%%%%%%%%%%%%%%%%%%%%%%%

The hints of Lepton Flavor Universality (LFU) violation in charged-current semi-leptonic $b\to c \ell \nu$
decays~\cite{Lees:2013uzd,Aaij:2015yra,Hirose:2016wfn,Aaij:2017deq,Abdesselam:2019dgh}, as well as in  $b\to s \ell\ell$ transitions~\cite{Aaij:2014ora,Aaij:2017vbb,Aaij:2019wad,Prim:2019wem}, represent  a very intriguing 
phenomenon and a fascinating challenge. 
Recent data confirm numerous discrepancies from the Standard Model (SM) predictions in both these sectors.
Despite the fact that there is not a single measurement with a high statistical significance, and that recent 
data have slightly decreased the overall significance of the anomalies, the global picture is still extremely interesting:
the internal consistency of available data is remarkable and, once combined, the significance of the LFU violating observables
exceeds $3.7\sigma$ in $b\to s \ell\ell$ and  $3.1\sigma$  in $b\to c \ell \nu$. 
A  common origin of the two sets of anomalies is not obvious, but it is a very appealing  possibility 
from the theoretical point of view. If confirmed as clear signals of physics beyond the Standard Model (SM),
the two anomalies combined would point to non-trivial dynamics at the TeV scale,
possibly linked to a solution of the SM flavor puzzle.

The initial efforts to address both sets of anomalies in terms of beyond the SM (BSM) physics  
have been focused on Effective Field Theory (EFT)  approaches 
(see~\cite{Bhattacharya:2014wla, Alonso:2015sja, Greljo:2015mma, Calibbi:2015kma} for the early attempts).
However, the importance of complementing EFT 
approaches with appropriate simplified models with new heavy mediators was soon realized~\cite{Greljo:2015mma},
and in this context leptoquark models played a key role~\cite{Hiller:2014yaa,Ghosh:2014awa,Varzielas:2015iva,Fajfer:2015ycq,Barbieri:2015yvd}.
Explicit heavy mediators are essential to address the compatibility of this class of SM extensions with other low-energy constraints and with high-$p_T$ data.
Due to the relatively low scale of new physics hinted at by the charged-current anomalies, high-$p_T$
constraints are indeed quite relevant~\cite{Dorsner:2016wpm,Faroughy:2016osc,Greljo:2017vvb,Diaz:2017lit,Hiller:2018wbv,Bansal:2018eha,Schmaltz:2018nls,Greljo:2018tzh,Baker:2019sli}.
Given the success of some EFT approaches and simplified models in describing available data, 
the attention has shifted recently towards the development of 
 more complete (and more complex) models with a consistent  ultraviolet (UV)  behavior
 (see in particular~\cite{DiLuzio:2017vat, Assad:2017iib, Calibbi:2017qbu,  Bordone:2017bld,  Bordone:2018nbg, Barbieri:2017tuq, DiLuzio:2018zxy, Marzocca:2018wcf,  Becirevic:2018afm,Greljo:2018tuh,Blanke:2018sro,Fornal:2018dqn,Altmannshofer:2017poe, Trifinopoulos:2018rna,Faber:2018qon}).

Already in early attempts~\cite{Alonso:2015sja,Barbieri:2015yvd}, 
the $U_1\sim(\mathbf{3},\mathbf{1})_{2/3}$ vector leptoquark, 
coupled mainly to third-generation fermions, emerged as an excellent
mediator for the explanation of both sets of anomalies. The effectiveness of this state as 
single-mediator accounting for all available low-energy data has been established in~\cite{Buttazzo:2017ixm}. 
However, the analysis of Ref.~\cite{Buttazzo:2017ixm}, as most other phenomenological analyses of the $U_1$ leptoquark
in $B$ physics (see in particular \cite{Bhattacharya:2016mcc,Angelescu:2018tyl,Kumar:2018kmr,Aebischer:2019mlg}), 
is based on the simplifying hypothesis of vanishing $U_1$ couplings to right-handed (RH) SM fermions. 
This hypothesis is motivated by the absence of clear indications of non-standard RH currents in present data, and by 
the sake of minimality, but it does not have a strong theoretical justification. Indeed 
the quantum numbers of the $U_1$ allow for RH couplings at the renormalizable level, 
and in motivated UV completions such couplings naturally appear~\cite{Bordone:2017bld,Bordone:2018nbg}.

The first goal of the present paper is the generalization of existing EFT/simplified-model 
studies on the $U_1$ impact in low-energy observables,
taking into account non-vanishing RH couplings (mainly to the third generation). 
As pointed out  first in~\cite{Bordone:2017bld,Bordone:2018nbg}
in the context of a specific UV completion, and as we show in more general terms below, such couplings lead to 
a series of interesting modifications in the low-energy phenomenology compared to the pure left-handed case.
We also update the analysis taking into account recent results on semileptonic $B$-meson
decays. New data by both LHCb~\cite{CERN-EP-2019-043} and 
Belle~\cite{Belle:MoriondRK} have not changed the overall 
significance of the anomalies in $b\to s \ell\ell$~\cite{Alguero:2019ptt,Straub:Moriond,Ciuchini:2019usw,Datta:2019zca}, while preliminary 
data from Belle~\cite{Belle:Moriond} have slightly decreased the significance in $b\to c \ell \nu$. However, as anticipated, 
the overall significance of the anomalies remains very high and the possibility of a combined 
explanation has become even  more consistent from a theoretical point of view,
due to the reduced tension with high-$p_T$ data and other low-energy observables.

The second goal of this paper is to assess whether the conditions necessary for a successful low-energy fit to present data,
compatible with high-$p_T$ constraints~\cite{Baker:2019sli}, 
can be achieved in the context of a consistent UV completion of the simplified model.
Here the main difficulty is to achieve a sizable 2-3 family mixing for the $U_1$ without introducing excessively large contributions 
to $\Delta F=2$ observables from other mediators required by the consistency of the theory.
As we show, this can be achieved by means of a simple extension of the scalar sector of the model 
originally proposed in~\cite{Bordone:2017bld} (see also~\cite{Bordone:2018nbg,Greljo:2018tuh}).

We provide a detailed implementation of the $U_1$ leptoquark in a renormalizable model based on the (flavor non-universal) 
gauge group $SU(4)_{3} \times SU(3)_{1+2} \times SU(2)_L \times U(1)^\prime$, which in turn 
can be embedded in PS$^3$~\cite{Bordone:2017bld}. In this context, we complement the simplified-model analysis
by including one-loop contributions to low-energy observables (most notably  $\Delta F=2$ amplitudes and dipole operators) 
which can be reliably computed only
within a UV-complete framework.

The paper is organized as follows. In Section~\ref{sec:LQDyn} we present the simplified-model analysis: we introduce the  
Lagrangian describing the $U_1$ couplings to SM fermions, and analyze its low-energy limit.
We discuss all the observables insensitive to the UV completion (Section~\ref{ssec:obs}), which are later used to fit  
low-energy data (Section~\ref{ssec:fit}). We finally comment 
 on the high-$p_T$ constraints  (Section~\ref{ssec:highpT}). 
 The UV-complete model is presented and discussed in Section~\ref{sec:UV}: on the model-building side 
 we pay particular attention to the flavor structure of the model (Section~\ref{sec:mixing}); 
 on the phenomenological side we present 
 complete expressions for the UV-dependent (loop-induced) 
 observables, which were omitted in the low-energy fit (Section~\ref{sec:loops}).
 The results are summarized in Section~\ref{sec:conclusions}.

%%%%%%%%%%%%%%%%%%%%%%%%%%%%%%%%%%%%%%%%%%%%%%%%%%%%%%%%%%%%%%%%%%
\section{The simplified $U_1$ model and its phenomenology}\label{sec:LQDyn}
%%%%%%%%%%%%%%%%%%%%%%%%%%%%%%%%%%%%%%%%%%%%%%%%%%%%%%%%%%%%%%%%%%

\subsection{Effective interactions of the $U_1$ to SM fields}
We consider the most general Lagrangian for the $SU(2)_L$-singlet vector leptoquark, $U_{1}^{\mu} \sim (\mathbf{3},\mathbf{1})_{2/3} $, coupled to both left- and right-handed SM fields
\begin{align}\label{eq:LQLag}
\begin{aligned}
\mathcal{L}_U=&-\frac{1}{2}\,U_{1\,\mu\nu}^\dagger\, U_1^{\mu\nu}+M_{U}^2\,U_{1\,\mu}^\dagger\, U_1^{\mu} -ig_c(1-\kappa_c)\,U_{1\,\mu}^\dagger\,T^a\,U_{1\,\nu}\,G^{a\,\mu\nu} \\
&-i \frac{2}{3} g_{Y}  (1-\kappa_Y)\,U_{1\,\mu}^\dagger\,\,U_{1\,\nu}\,B^{\mu\nu}+(U_1^\mu J_\mu + \mathrm{h.c.})\,,
\end{aligned}
\end{align}
where $U_{1\, \mu \nu} = D_{\mu} U_{1\,\nu} - D_{\nu} U_{1\,\mu}\,,$ with $D_{\mu} = \partial_{\mu} - i g_c\, G_{\mu}^{a}T^{a} - i \frac23 g_{Y}  B_{\mu}$. Here $G_\mu^a$ ($a=1,\dots,8$) and $B_\mu$ denote the SM $SU(3)_c$ and $U(1)_Y$ gauge bosons, with $g_c$ and $g_Y$ gauge couplings respectively, and $T^a$ are the $SU(3)_c$ generators. In models in which the vector leptoquark has a gauge origin, $\kappa_c=\kappa_Y=0$, while this is not necessarily the case for models in which the $U_1	$ arises as a bound state from a strongly-coupled sector. The fermion current reads\footnote{We ignore possible couplings to right-handed neutrinos. Such particles, if present, are assumed to be heavy enough such that they do not to play any role in low-energy observables. Vector leptoquark solutions of the $B$-physics anomalies involving right-handed neutrinos, light enough to fake the SM ones, have been discussed in~\cite{Azatov:2018kzb,Robinson:2018gza}.}
\begin{align}
J_\mu=\frac{g_U}{\sqrt{2}}  \left[ \beta_{L}^{i\alpha}\,(\bar q_{L}^{\,i} \gamma_{\mu}  \ell_{L}^{\alpha})  +     \beta_{R}^{i \alpha }\,(\bar d_{R}^{\,i}\gamma_{\mu}   e_{R}^{\alpha})\right] \,.
\end{align} 
Here the couplings $\beta_{L}$ and $\beta_{R}$ are complex $3 \times 3$ matrices in flavor space. Without loss of generality, we adopt the down-quark and charged-lepton mass eigenstate basis for the $SU(2)_L$ multiplets, i.e.
\begin{align}\label{eq:DownBasis}
q_L^i=
\begin{pmatrix}
V_{ji}^*\,u^j_L\\
d_L^i
\end{pmatrix}
\,,\qquad\qquad 
\ell_L^i=
\begin{pmatrix}
\nu^i_L\\
e_L^i
\end{pmatrix}
\,.
\end{align}
In this flavor basis, we assume the following structure for the $\beta_{L}$ and $\beta_{R}$ couplings
\begin{equation}\label{eq:minCoup}
\begin{aligned}
\beta_{L} = 
\begin{pmatrix} 
0 & 0 & \beta_L^{d\tau} \\[5pt]
0 & \beta_{L}^{s \mu} &  \beta_{L}^{s \tau}\\[5pt]
0 & \beta_{L}^{b \mu} &  1
\end{pmatrix}
\,,\qquad\qquad 
\beta_{R} = 
\begin{pmatrix} 
0 & 0 & 0 \\[5pt]
0 & 0 & 0\\[5pt]
0 & 0 &  \beta_{R}^{b \tau} 
\end{pmatrix}
\,,
\end{aligned}
\end{equation}
where the normalization of $g_{U}$ is chosen such that $\beta_{L}^{b \tau} =1$. 
The assumed structure contains the minimal set of couplings directly connected to a combined explanation of the 
$B$-physics anomalies. The null entries in Eq.~(\ref{eq:minCoup}) should be understood as small terms which have a negligible 
impact in the observables we analyze. We discuss later on the implications of this requirement on the values of $\beta_L^{d\mu}$ and $\beta_R^{b \mu}$. 
Under the assumption of a natural (CKM-like) flavor structure, the entries in~\eqref{eq:minCoup} are expected to follow the relations
 $\beta_{L}^{d \tau},~\beta_{L}^{s \mu}  \ll  \beta_{L}^{s \tau},~\beta_{L}^{b \mu}  \ll 1$, and   $\beta_{R}^{b \tau} = \mathcal{O}(1)$. 
 As we shall see, 
the hierarchy of the $\beta$'s is well compatible and, at least in some cases, it emerges from the fit to low-energy data. 
The only parameter we force to be small {\em a priori} in the phenomenological analysis is $\beta_L^{s\tau}$, which 
is largely unconstrained in the simplified model (using only low-energy data) and 
plays a key role in setting the overall mass scale for the $U_1$. We set the upper limit  
$|\beta_L^{s\tau} | \leq  0.25$~(see Section~\ref{ssec:fit}). 
 In Section~\ref{sec:UV} we show how this hierarchical structure of the $\beta$'s is naturally enforced 
 in the proposed UV completion.

By integrating out the vector leptoquark at tree level, we obtain the following high-scale ($\mu \sim M_U$) 
effective Lagrangian:
\begin{equation}
\begin{aligned}
\mathcal{L}_{\mathrm{eff}}  = - \frac{2  C_{U} }{v^{2}}  & \left[   -2 \, (\beta_L^{i \alpha})^*  \beta_{R }^{ l \beta} (\bar \ell_{L}^{\alpha}e_{R}^{\beta}  ) (\bar  d_{R}^{l}  q_{L}^{i} ) + \mathrm{h.c.} + \beta_{R}^{i \alpha}  (\beta_R^{ l \beta})^* (\bar e_{R}^{\beta}  \gamma_{\mu} e_{R}^{\alpha}) (\bar d_{R}^{i} \gamma^{\mu}  d_{R}^{l}) 
 \right.\\
 & \left.  +\frac12 \beta_{L}^{i \alpha}  (\beta_{L}^{ l \beta})^* (\bar \ell_{L}^{\beta}  \gamma_{\mu} \ell^{\alpha}_{L})(\bar q^{i}_{L}  \gamma^{\mu}q_{L}^{l} )  +   \frac12 \beta_{L}^{i \alpha}  (\beta_L^{ l \beta})^*  (\bar \ell_{L}^{\beta} \sigma^{a}   \gamma_{\mu} \ell^{\alpha}_{L})(\bar q^{i}_{L}  \sigma^{a} \gamma^{\mu}q_{L}^{l} )  \right] \,,
\end{aligned}
\label{eq:eft_lag}
\end{equation}
where $C_{U} \equiv g_{U}^{2} v^{2}/(4 M_{U}^{2})$ and $v=(\sqrt{2}\, G_F)^{-1/2}\approx246$~GeV is the SM Higgs vacuum expectation value (vev).

%%%%%%%%%%%%%%%%%%%%%%%%%%%%%%%%%
\subsection{The relevant low-energy observables}\label{ssec:obs}
%%%%%%%%%%%%%%%%%%%%%%%%%%%%%%%%%

Since the Lagrangian in~\eqref{eq:LQLag} is not renormalizable, only a limited set of low-energy observables can be reliably estimated 
in this setup. These are observables where the four-fermion interactions in (\ref{eq:eft_lag}) contribute at the tree level,
or where they induce (log- or Yukawa-enhanced) loop contributions which are largely insensitive to the UV completion.
In this section and in the corresponding low-energy fit we consider only such class of observables. 
The discussion of the UV-dependent one-loop contributions is postponed to Section~\ref{sec:UV}.

Taking into account these considerations,  the most relevant low-energy observables 
can be classified as follows: 
\begin{itemize}
\item[i)] {\bf $b\to c(u)\tau\nu$.} Sizable tree-level contributions arise in the LFU ratios $R_{D}$ and $R_{D^{*}}$. 
Due to the presence of the right-handed coupling, $\beta_R^{b\tau}$, the  usual V-A contribution is supplemented by a (large) scalar contribution, yielding the following approximate expressions for $R_{D}$ and $R_{D^{*}}$ (see~\cite{Feruglio:2018fxo,Fajfer:2012vx} for the hadronic matrix elements of the scalar contribution)
\begin{align}
R_{D} &\approx R_{D}^{\mathrm{SM}}\left[1+ 2 C_{U}\, \mathrm{Re} \left\lbrace \left( 1-1.5\, \eta_{S}\, (\beta_R^{b \tau})^* \right)\left(1+ \frac{V_{cs}}{V_{cb}} \beta_L^{s \tau} + \frac{V_{cd}}{V_{cb}} \beta_L^{d \tau}\right) \right\rbrace \right]\,, \label{eq:RD}\\
 R_{D^{*}} &\approx  R_{D^{*}}^{\mathrm{SM}} \left[1+ 2 C_{U} \,\mathrm{Re}\left \lbrace \left(  1 -0.14 \,\eta_{S} \,(\beta_R^{b \tau})^* \right)\left( 1  + \frac{V_{cs}}{V_{cb}} \beta_L^{s \tau} + \frac{V_{cd}}{V_{cb}} \beta_L^{d \tau}\right) \right\rbrace \right]\,, \label{eq:RDs}
 \end{align}
where $\eta_S\approx1.8$ accounts for the running of the scalar operator from $M_U=4$~TeV to $m_b$~\cite{Gonzalez-Alonso:2017iyc,Celis:2017hod,Aebischer:2018bkb}. 
Interestingly, due to the scalar contribution, we obtain a significantly different scaling of the NP effect in $R_{D}$ and in $R_{D^{*}}$, depending on the value of $\beta_R^{b\tau}$ (see Figure~\ref{fig:proyections}). 
 
Additionally, because of the chiral enhancement of the scalar contribution, large NP effects are expected in $\mathcal{B}( B_c \to \tau\nu)$ 
 \begin{align}\label{eq:Bc2taunu}
\begin{aligned}
\mathcal{B}( B_c \to \tau\nu) &=\frac{\tau_{B_c}\,m_{B_c}f_{B_c}^2 G_F^2|V_{cb}|^2}{8\pi}\,m_\tau^2\left(1-\frac{m_\tau^2}{m_{B_c}^2}\right)^2 \\
&\quad\times\left|1+ C_U\left (1 -(\beta_R^{b \tau})^*\frac{2\,\eta_S\,m_{B_c}^2}{m_\tau(m_b + m_c)}\right)\left(1+\frac{V_{cs}}{V_{cb}}\beta_L^{s \tau}+\frac{V_{cd}}{V_{cb}}\beta_L^{d \tau}\right) \right|^2 \,.  
\end{aligned}
\end{align}
The most stringent bounds on this observable are obtained from LEP data from which the authors of~\cite{Akeroyd:2017mhr} obtain $\mathcal{B}( B_c \to \tau\nu)\lesssim10\%$ (see also~\cite{Alonso:2016oyd}). 

Concerning the $b\to u$ transitions, the only measured observable in this category is $\mathcal{B}( B \to \tau\nu)$, for which we obtain the following expression
\begin{align}
\mathcal{B}( B \to \tau\nu) &= \mathcal{B}( B \to \tau\nu)_{\rm SM} \left|1+ C_U\left (1 -(\beta_R^{b \tau})^*\frac{2\,\eta_S\,m_B^2}{m_\tau(m_b + m_u)}\right)\left(1+\frac{V_{us}}{V_{ub}}\beta_L^{s \tau}+\frac{V_{ud}}{V_{ub}}\beta_L^{d \tau}\right) \right|^2 \,.  \label{eq:B2taunu} 
\end{align}
Also here, we expect large NP effects due to the chirally enhanced scalar contribution. However, in this case the connection with $R_{D^{(*)}}$ is less robust due to the possible sizable contribution from $\beta_L^{d\tau}$, which now receives a larger CKM-enhancement than the one from $\beta_L^{s\tau}$. 

\begin{figure}[!t]
\centering
\includegraphics[width=0.45\textwidth]{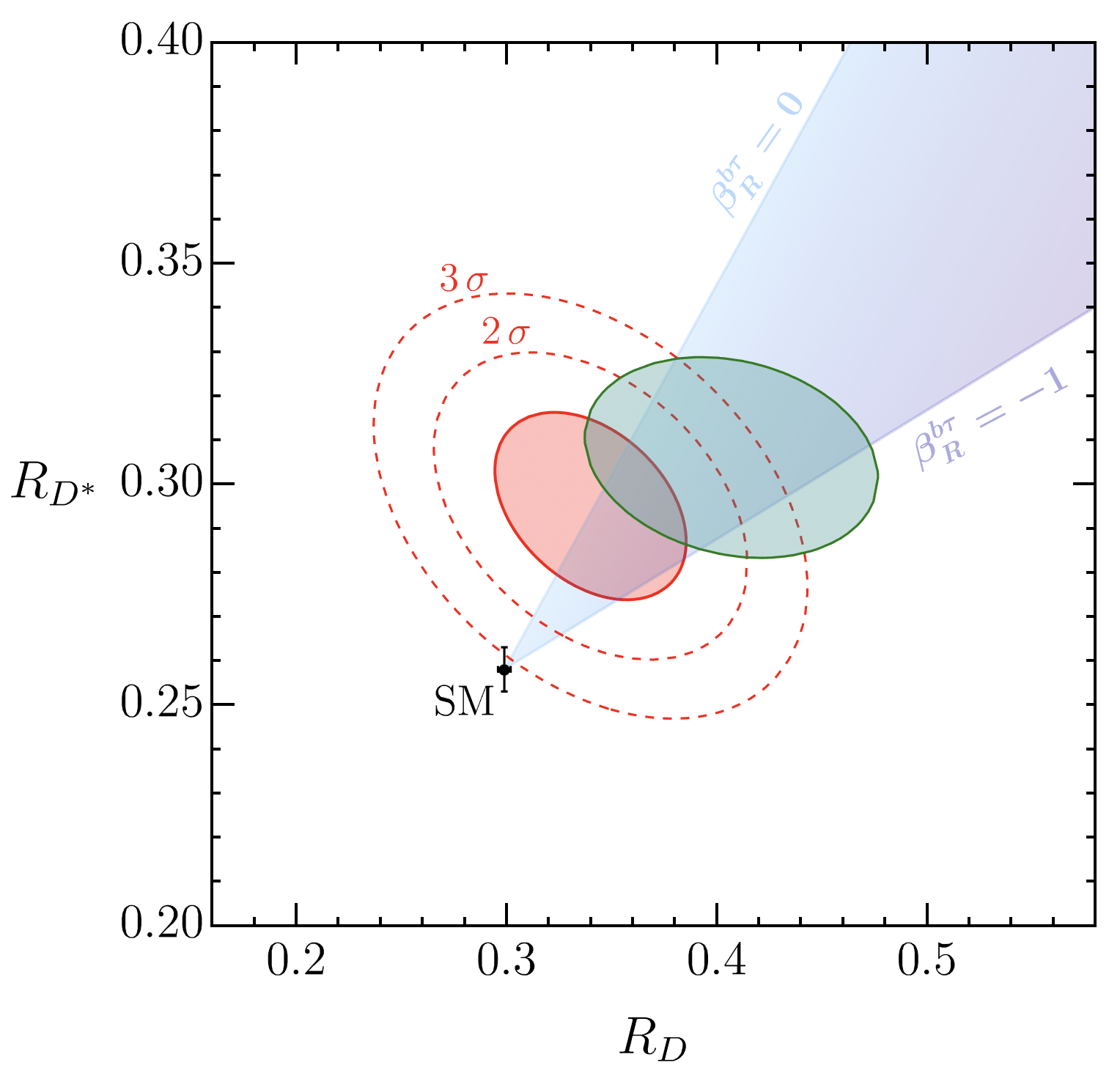} 
\caption{$SU(2)_L$-singlet vector leptoquark NP projections for $R_D$ and $R_{D^*}$ as a function of $\beta_R^{b\tau}$, together with the latest experimental world average (in red) and the SM prediction. For illustration, we also show the experimental $1\sigma$ HFLAV combination~\cite{Amhis:2016xyh}  (in green), previous to the inclusion of preliminary Belle data announced in~\cite{Belle:Moriond}.}
\label{fig:proyections}
\end{figure}

\item[ii)] {\bf $b\to s\ell\ell$.} The vector leptoquark  yields potentially large contributions to $b\to s\ell\ell$ transitions, both at tree level and at one loop. Given our assumption of  vanishing couplings to electrons ($\beta_{L,R}^{qe}=0$, for any $q$), tree-level contributions affect only $b\to s \mu \mu$ and $b\to s \tau \tau$ transitions along the direction $\Delta \mathcal{C}_{9}^{\ell \ell} =  - \Delta \mathcal{C}_{10}^{\ell \ell}$ with $\ell=\mu,\tau$ (see Appendix~\ref{app:EffHam} for the Wilson coefficient definitions)
\begin{align}\label{eq:C9NU} 
\Delta \mathcal{C}_{9}^{\ell \ell} &=  - \Delta \mathcal{C}_{10}^{\ell \ell}  = - \frac{2 \pi}{ \alpha V_{tb} V_{ts}^{\ast}} \, C_{U} \beta_L^{s \ell} (\beta_L^{b \ell})^*\,. 
\end{align} 

Being lepton non-universal, these Wilson coefficient modifications affect the $R_{K^{(*)}}$ ratios in the following way~\cite{Celis:2017doq,Capdevila:2017bsm}
\begin{align}
\begin{aligned}
\Delta R_K&\equiv R_K^{[1,6]\, \mathrm{GeV}^2}-1\approx 0.46  \, \Delta \mathcal{C}_{9}^{\mu \mu}\,,\\
\Delta R_{K^*}&\equiv R_{K^*}^{[1.1,6]\, \mathrm{GeV}^2}-1\approx 0.47\, \Delta \mathcal{C}_{9}^{\mu \mu}\,.
\end{aligned}
\end{align}

As discussed in~\cite{Crivellin:2018yvo},  a large $\beta_L^{s\tau}$ coupling can also yield a sizable lepton-universal contribution to $b \to s \ell \ell$ transitions in the $\Delta \mathcal{C}_{9}$ direction via a (log-enhanced) photon penguin. Since the 
dominant contribution is given by the log-enhanced piece, it can be unambiguously recovered from the corresponding 
EFT computation which gives
\begin{align}\label{eq:C9U}
\begin{aligned}
\Delta \mathcal{C}_{9}^U &\approx -\frac{1}{V_{tb}V_{ts}^*}\,\frac{2}{3}\,C_U\sum_{\ell=e,\,\mu,\,\tau} 
\beta_L^{s\ell}(\beta_L^{b\ell})^*\,\log(m^2_b/M_U^2)  \\
&\approx -\frac{1}{V_{tb}V_{ts}^*}\,\frac{2}{3}\,C_U 
\beta_L^{s\tau}(\beta_L^{b\tau})^*\,\log(m^2_b/M_U^2)~.
\end{aligned}
\end{align}
For non-zero $\beta_R^{b\mu}$, scalar-current contributions are generated in $b\to s\mu\mu$ transitions.
As shown in~\cite{Bordone:2018nbg}, a stringent bound on $\beta_R^{b\mu}$ follows from $B_s \to \mu^+\mu^-$, since the scalar-current contribution is chirally enhanced. 
Fixing the other parameters to fit the current central value of $R_{K^{(*)}}$, present data imply $|\beta_R^{b\mu}| \lsim 0.02\,|\beta_L^{b\mu}|$. Once this condition is imposed, the effect of 
$\beta_R^{b\mu}$ on other $b\to s\mu\mu$ observables is negligible. We can thus directly compare the 
corrections to $\mathcal{C}_{9, 10}^{\mu\mu}$ in~\eqref{eq:C9NU} and $\mathcal{C}_9^U$ in~\eqref{eq:C9U}
with the global fits of these Wilson coefficients reported in~\cite{Alguero:2019ptt,Straub:Moriond} (see also~\cite{Alguero:2018nvb,Capdevila:2017bsm,Descotes-Genon:2015uva} for details on the fit methodology followed in~\cite{Alguero:2019ptt}). 

On the other hand, scalar currents are necessarily present in $b\to s\tau\tau$ transitions if $\beta_R^{b\tau} =\mathcal{O}(1)$.
The most interesting observable in this respect is 
\begin{align}
\mathcal{B}(B_{s} \to \tau^{+} \tau^{-}) &= \mathcal{B}(B_{s} \to \tau^{+} \tau^{-})_{\rm SM} \left[\left|1+\frac{2\pi}{\alpha V_{tb}V_{ts}^*}\frac{C_U}{C_{10}^{\rm SM}}\, \beta_L^{s \tau}\left(1  - \frac{\eta_{S}\, m_{B_{s}}^{2}  }{m_{\tau}(m_{s} + m_{b})}\,  (\beta_R^{b \tau})^* \right)\right|^{2} \right. \nonumber \\
&\quad\left.+\left(1-\frac{4m_\tau^2}{m_{B_s}^2}\right)\left|\frac{2\pi}{\alpha V_{tb}V_{ts}^*}\frac{C_U}{C_{10}^{\rm SM}} \frac{\eta_{S}\, m_{B_{s}}^{2}  }{m_{\tau}(m_{s} + m_{b})}\, \beta_L^{s \tau} (\beta_R^{b \tau})^* \right|^{2} \right] \,. \label{eq:Bs2tautau}
\end{align}
A milder, but still sizable, chiral enhancement occurs in $\mathcal{B}(B \to K \tau^+\tau^-)$, which can be large and within the reach of future experiments, especially for $\beta_R^{b\tau} =\mathcal{O}(1)$. Using the hadronic form factors in~\cite{Bouchard:2013pna} we find
\begin{align}\label{eq:B2Ktautau} 
\begin{aligned}
\mathcal{B}(B \to K \tau^{+} \tau^{-})  &\approx 1.5\cdot10^{-7}+10^{-3}\, C_U\,(1.4\,\mathrm{Re}\lbrace\beta_L^{s\tau}\rbrace-3.3\,\mathrm{Re}\lbrace\beta_L^{s\tau}\beta_R^{b\tau\,*}\rbrace) \\
&\quad+ C_U^2\,|\beta_L^{s\tau}|^2\, (3.5 - 16.4\, \mathrm{Re}\lbrace\beta_R^{b\tau}\rbrace+ 95.0\, |\beta_R^{b\tau}|^2)\,.
\end{aligned}
\end{align}

An interesting feature of the vector-leptoquark solution is the absence of tree-level contributions to $b\to s\nu\nu$ observables, letting this setup easily pass the current constraints from $B\to K^{(*)}\nu\nu$.

\item[iii)] {\bf Dipoles.} For $\beta_R^{b \tau} \neq 0$, the presence of both left- and right-handed leptoquark couplings gives rise to contributions to the radiative LFV decay $\tau\to\mu\gamma$ that are $m_b$-enhanced. Taking $\kappa_Y=0$, the $m_b$-enhanced piece is finite and can be unambiguously computed already in the dynamical model. We find
\begin{align}\label{eq:tau2mugamma}
\mathcal{B}(\tau \to \mu \gamma) & \approx  \frac{1}{\Gamma_{\tau}} \frac{\alpha}{64 \pi^{4}} \frac{m_{\tau}^3m_b^2 }{v^{4}}\, C_{U}^{2} | \beta_R^{b \tau} (\beta_L^{b \mu})^*|^{2} \,.
\end{align}
Analogous loop effects in the $b\to s\gamma(g)$ transitions are more sensitive to the specifics of the UV completion.
Indeed the contribution proportional to the internal mass in the $U_1$-mediated amplitude leads to a $\mathcal{O}(m_\tau^2/m_b^2)$
suppression, rather than an enhancement, compared to the one proportional to the external mass. This latter contribution is sensitive to the details of the UV completion and cannot be reliably estimated. 
We thus postpone their discussion to Section~\ref{sec:UV}.

\item[iv)] {\bf LFV observables}. The vector leptoquark can also yield sizable tree-level contributions to semileptonic LFV transitions. The most interesting observables are those involving the $b \to s \tau \mu$ transition. One of the observables in this category for which experimental limits are available is $B^+ \to K^+ \tau \mu$. The simplified expressions are given by~\cite{Bordone:2018nbg}
\begin{align}\label{eq:B2Ktaumu} 
\mathcal{B}(B^{+} \to K^{+} \tau^{+} \mu^{-})  &\approx C_U^2\, |\beta_L^{s\mu}|^{2} \left(8.3 + 155.2\, |\beta_R^{b \tau}|^{2} - 42.3\, \mathrm{Re} \lbrace \beta_R^{b \tau}\rbrace \right)\,,\\
\mathcal{B}(B^{+} \to K^{+} \tau^{-} \mu^{+})  &\approx 8.3\,C_U^2\, |\beta_L^{b\mu}(\beta_L^{s\tau})^*|^{2} \,.
\end{align}
Note that, for large values of $\beta_R^{b \tau}$, the $\tau^{+} \mu^{-}$ channel is expected to yield the dominant NP contribution, provided the other couplings follow the natural flavor scaling discussed in Section~\ref{sec:LQDyn}.

As in $B_{s} \to \tau \tau$, the NP effect in $B_{s} \to \tau \mu$ is chirally enhanced for $\beta_R^{b \tau} \neq 0$, making this observable of particular interest. Its expression reads
\begin{align}\label{eq:Bs2taumu}
&\mathcal{B}(B_{s} \to \tau^{-} \mu^{+}) = \frac{\tau_{B_s} m_{B_{s}}  f_{B_{s}}^{2} G_{F}^{2}}{8 \pi}\, m_{\tau}^{2} \left(1 - \frac{m_{\tau}^{2}}{m_{B_{s}}^{2}}\right)^{2}  C_{U}^{2}  \left| \beta_{L }^{s \mu} (\beta_L^{b \tau})^* - \frac{2\,\eta_{S}\, m_{B_{s}}^{2}}{m_{\tau}(m_{s} + m_{b})}\, \beta_L^{s \mu} (\beta_R^{b \tau})^*\right|^{2} \,. 
\end{align}
The LHCb Collaboration has recently performed the first measurement of this observable, setting the upper limit 
$\mathcal{B}(B_{s} \to \tau^{\pm} \mu^{\pm}) < 4.2\times 10^{-5}$ at 95\%~C.L.~\cite{Aaij:2019okb}, whose 
 implications  are discussed in the next section.

Another interesting LFV observable, relevant in the limit of large $\beta_L^{s \tau}$, is $\tau \to \mu \phi$ (see e.g.~\cite{Bhattacharya:2016mcc}).
Here we find
\begin{align}
\mathcal{B}( \tau \to \mu \phi)& = \frac{1}{\Gamma_{\tau}} \frac{ f_{\phi}^{2} \, G_{F}^{2} }{16 \pi } m_{\tau}^{3} \left(1-  \frac{m_\phi^2}{m_\tau^2}\right)^{2} \left(1 + 2 \frac{m_\phi^2}{m_\tau^2}\right)\, C_{U}^{2} \left|\beta_L^{s \tau}(\beta_L^{s \mu})^*\right|^{2} \,. \label{eq:tau2muphi} 
\end{align} 

\item[v)] {\bf LFU in $\tau$ decays}.
At the one-loop level,  the effective Lagrangian in~\eqref{eq:eft_lag} leads to modifications of the $Z$ and $W$ couplings to fermions
and, more generally,  to LFU breaking effects in purely leptonic charged-current transitions, 
as extensively discussed in ~\cite{Feruglio:2016gvd,Feruglio:2017rjo,Cornella:2018tfd}. The most constraining bounds arise from LFU tests in $\tau$ decays, in particular from the ratio  $g_{\tau}/g_{\mu}$. Using the results in~\cite{Bordone:2018nbg},
we can describe these effects via the following simplified expression
\begin{align}
 \left( \frac{g_{\tau}}{g_{\mu}} \right)_{\ell, \pi, K} &  \approx 1 -  0.08 \, C_{U}\,, \label{eq:LFU} 
\end{align}
where we have set $M_{U} = 4$~TeV in the evaluation of the leptoquark loop.

\item[vi)] {\bf $\Delta F=2$ observables}. Though important, loop contributions to $\Delta F=2$ transitions mediated by the vector leptoquark are divergent and cannot be reliably estimated without a UV completion. The discussion of these effects is therefore postponed to Section \ref{sec:loops}.
\end{itemize}

\begin{table}[ht]
\centering
\renewcommand{\arraystretch}{1.2} 
\begin{tabular}{c |c c |c|c }
Observable & Experiment  & Corr. & SM  & $U_1$ expression  \\
\hline 
$R_{D} $& \multirow{2}{*}{$\begin{matrix} 0.340(30) \\ 0.295(13) \end{matrix}$\cite{HFLAV:RDRDs}}  &  \multirow{2}{*}{$-0.37$}  & $0.299(3)$ \cite{Bigi:2016mdz,Bernlochner:2017jka,Jaiswal:2017rve}&  \eqref{eq:RD} \\
 $R_{D^{*}}$ &  & & $0.258(5)$ \cite{Bigi:2017jbd,Bernlochner:2017jka,Jaiswal:2017rve} & \eqref{eq:RDs} \\
$\mathcal{B}( B \to \tau \nu)$ &$1.09(24)\cdot 10^{-4}$ \cite{Tanabashi:2018oca} & $-$ &$0.812(54) \cdot 10^{-4}$ \cite{Bona:2017cxr} &\eqref{eq:B2taunu} \\
$\Delta \mathcal{C}_{9}^{\mu \mu}=-\Delta \mathcal{C}_{10}^{\mu \mu}$ &  \multirow{2}{*}{$\begin{matrix} -0.40\pm0.12 \\  -0.50\pm 0.38 \end{matrix}$~\cite{Alguero:2019ptt,Straub:Moriond}}  & \multirow{2}{*}{ $-0.5$} & $-$ &  \eqref{eq:C9NU} \\
$\Delta \mathcal{C}_{9}^{U}$ & & $$ & $-$ &  \eqref{eq:C9U} \\
$\mathcal{B}(B_{s} \to \tau^{+} \tau^{-})$ & $0.0(3.4) \cdot 10^{-3}$ \cite{Aaij:2017xqt} & $-$ &$7.73(49)\cdot 10^{-7}$ \cite{Bobeth:2013uxa} &\eqref{eq:Bs2tautau}\\
$\mathcal{B}(B^+ \to K^+ \tau^{+} \tau^{-})$ & $1.36(0.71) \cdot 10^{-3}$ \cite{TheBaBar:2016xwe} & $-$ &$1.5(0.2)\cdot 10^{-7}$ &\eqref{eq:B2Ktautau}\\
$\mathcal{B}(\tau \to \mu \gamma) $ & $0.0(3.0) \cdot 10^{-8}$ \cite{Amhis:2016xyh} & $-$ & $-$ &\eqref{eq:tau2mugamma}\\
$\mathcal{B}(B^{+} \to K^{+} \tau^{+} \mu^{-}) $ & $0.0(1.7) \cdot 10^{-5}$ \cite{Lees:2012zz} & $-$ & $-$ & \eqref{eq:B2Ktaumu} \\
$\mathcal{B}(B_s \to \tau^{\pm} \mu^{\mp}) $ & $0.0(2.1) \cdot 10^{-5}$ \cite{Aaij:2019okb} & $-$ & $-$ & \eqref{eq:Bs2taumu} \\
$\mathcal{B}( \tau \to \mu \phi)$ & $0.0(5.1) \cdot 10^{-8}$ \cite{Miyazaki:2011xe}& $-$ & $-$ &\eqref{eq:tau2muphi} \\
$ ( g_{\tau}/g_{\mu} )_{\ell, \pi, K} $ & $1.0000 \pm 0.0014$ \cite{Amhis:2016xyh} & $-$ & $1.$ &\eqref{eq:LFU} \\
\end{tabular}
\label{tab:fitobs}
\caption{List of observables included in the fit.
 The experimental values and SM predictions are also shown. The expressions of the observables in terms of the $U_1$ parameters are reported in Section~\ref{ssec:obs}.}
\end{table}

%%%%%%%%%%%%%%%%%%%%%%%%%%%%%%%%%
\subsection{Fit to low-energy data}\label{ssec:fit}
%%%%%%%%%%%%%%%%%%%%%%%%%%%%%%%%%
We are now ready to assess the phenomenological impact of the observables discussed in the previous section. In order to simplify the discussion we fix $\beta_R^{b \tau} = -1$. While solutions to the $B$-meson anomalies where $\beta_R^{b \tau}\neq-1$ are possible, and are even 
slightly favored by the latest data,\footnote{The global fits (considering only the low-energy observables in Sect.~2.2)
obtained with $\beta_R^{b \tau} = 0$ and $\beta_R^{b \tau} = -1$ differ by $\Delta \chi^2=1.7$, which is not statistically significant.}  the parameter $\beta_R^{b \tau}$ is not tightly 
constrained and we find it useful to fix it to  
$\beta_R^{b \tau} = -1$ for three  main reasons. First, we want  to stress the main differences of this scenario with respect to the often discussed solution in which $\beta_R^{b \tau} = 0$~\cite{Buttazzo:2017ixm,Angelescu:2018tyl,Straub:Moriond}. Second, this solution maximizes the NP contribution to $\Delta R_{D^{(*)}}$ (for a fixed value of $g_U/M_U$), allowing us to lift the NP mass spectrum, a very desirable feature in view of the tight high-$p_T$ constraints on TeV-scale mediators. Finally, as we show in Section~\ref{sec:UV}, one expects $|\beta_R^{b \tau}|\approx 1$ 
in  the explicit UV completions we are considering.

 \begin{figure}[!t]
\centering
\includegraphics[width=0.47\textwidth]{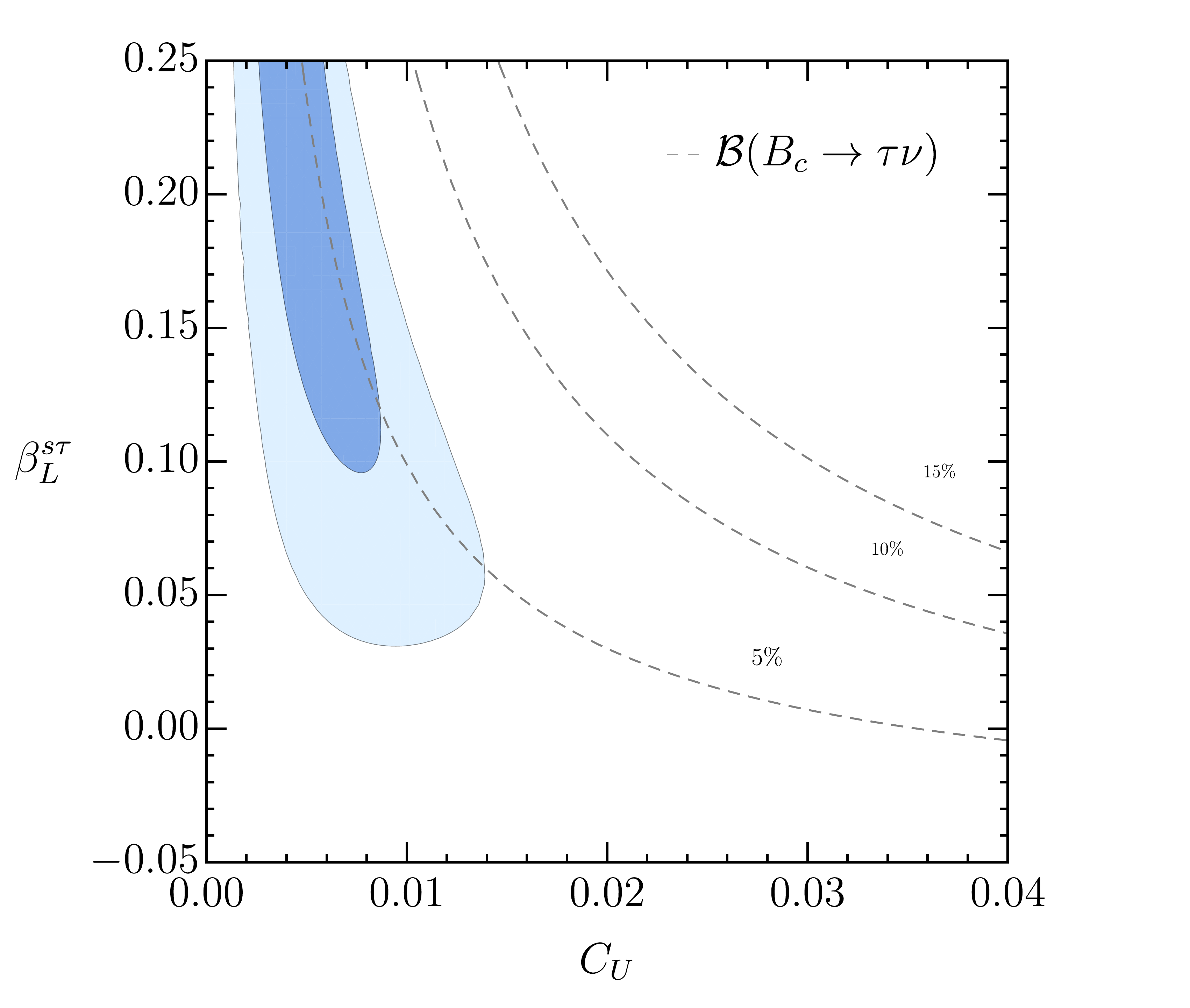} ~~ \includegraphics[width=0.47\textwidth]{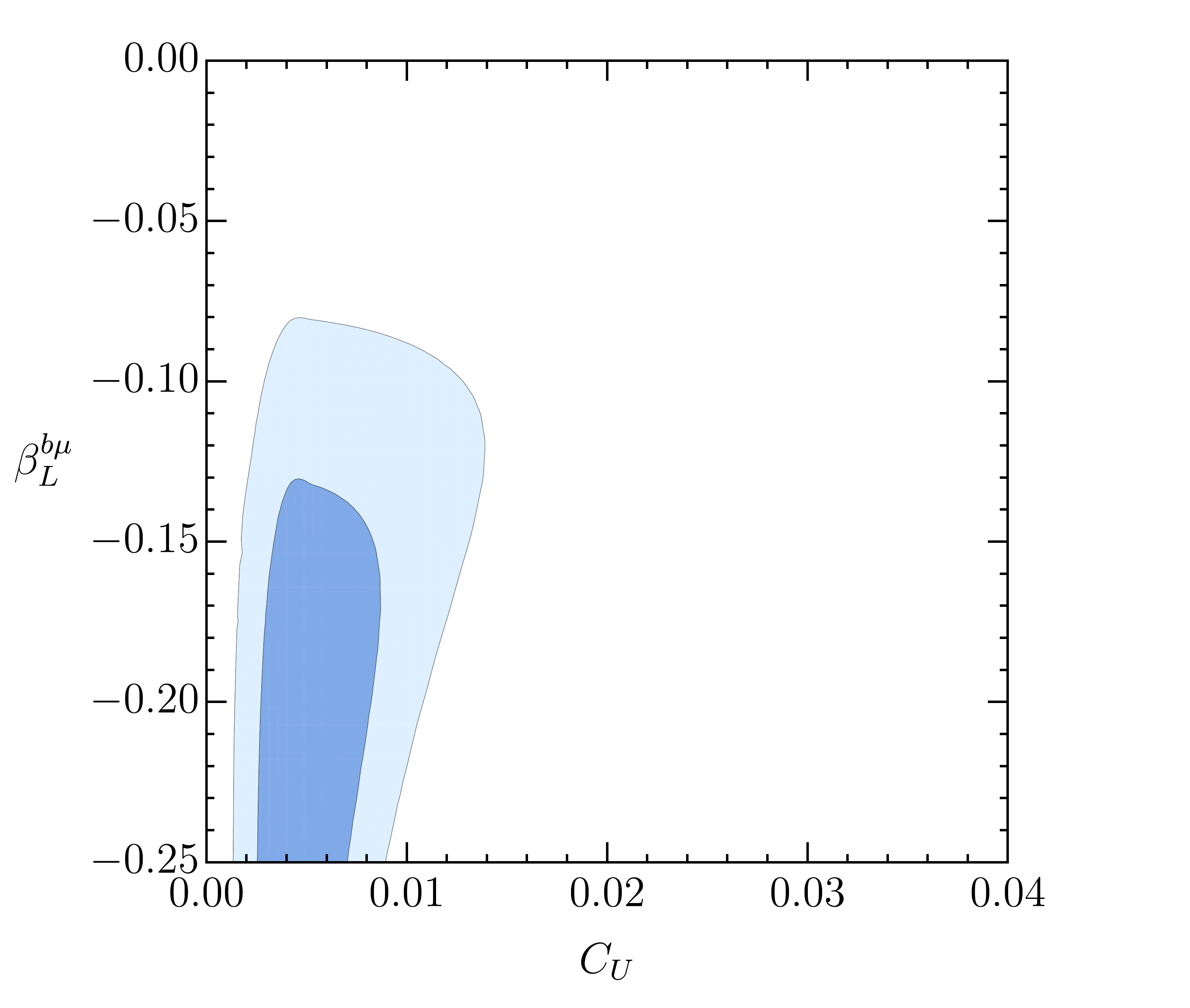} \\ 
\includegraphics[width=0.47\textwidth]{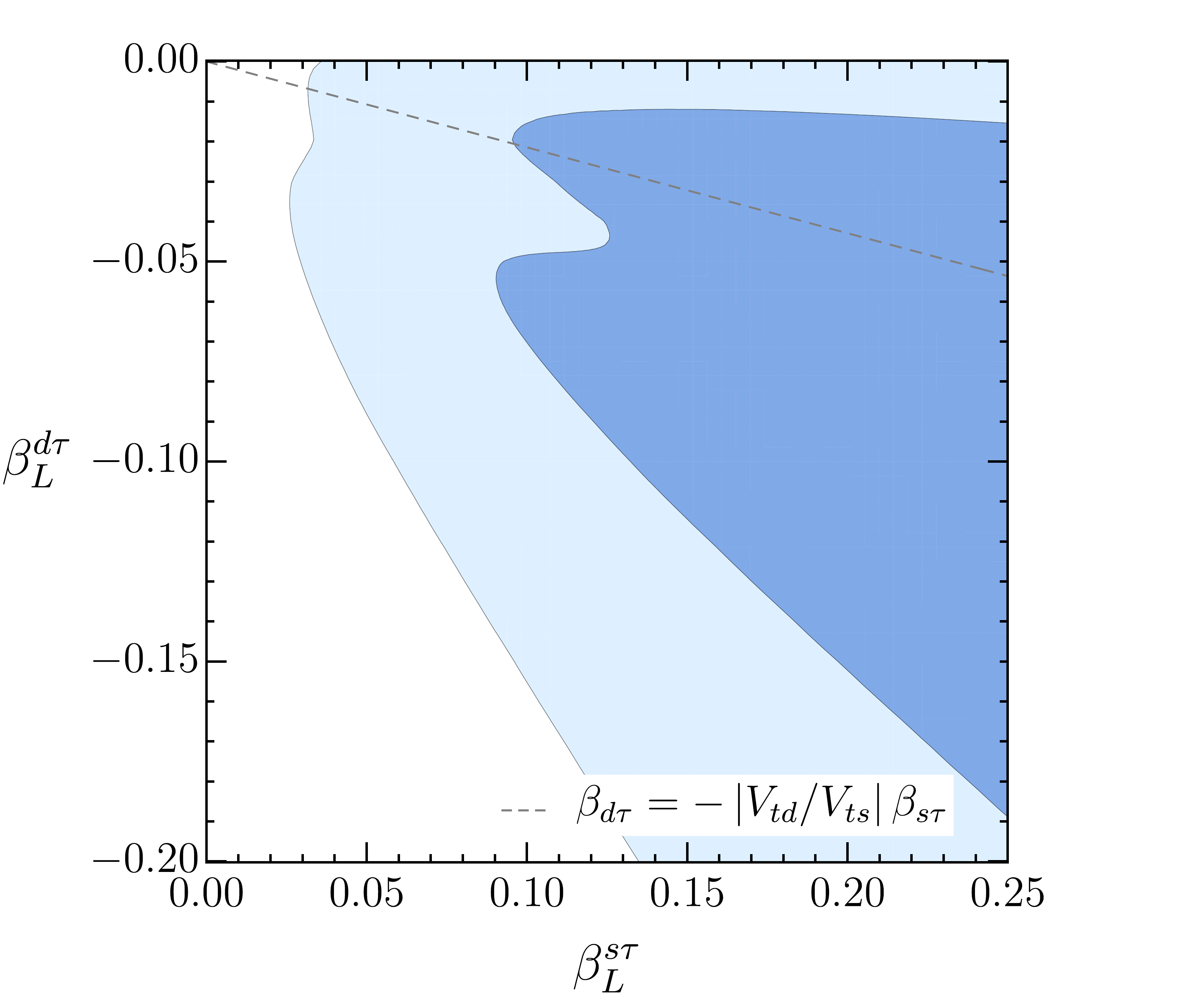} ~~ \includegraphics[width=0.47\textwidth]{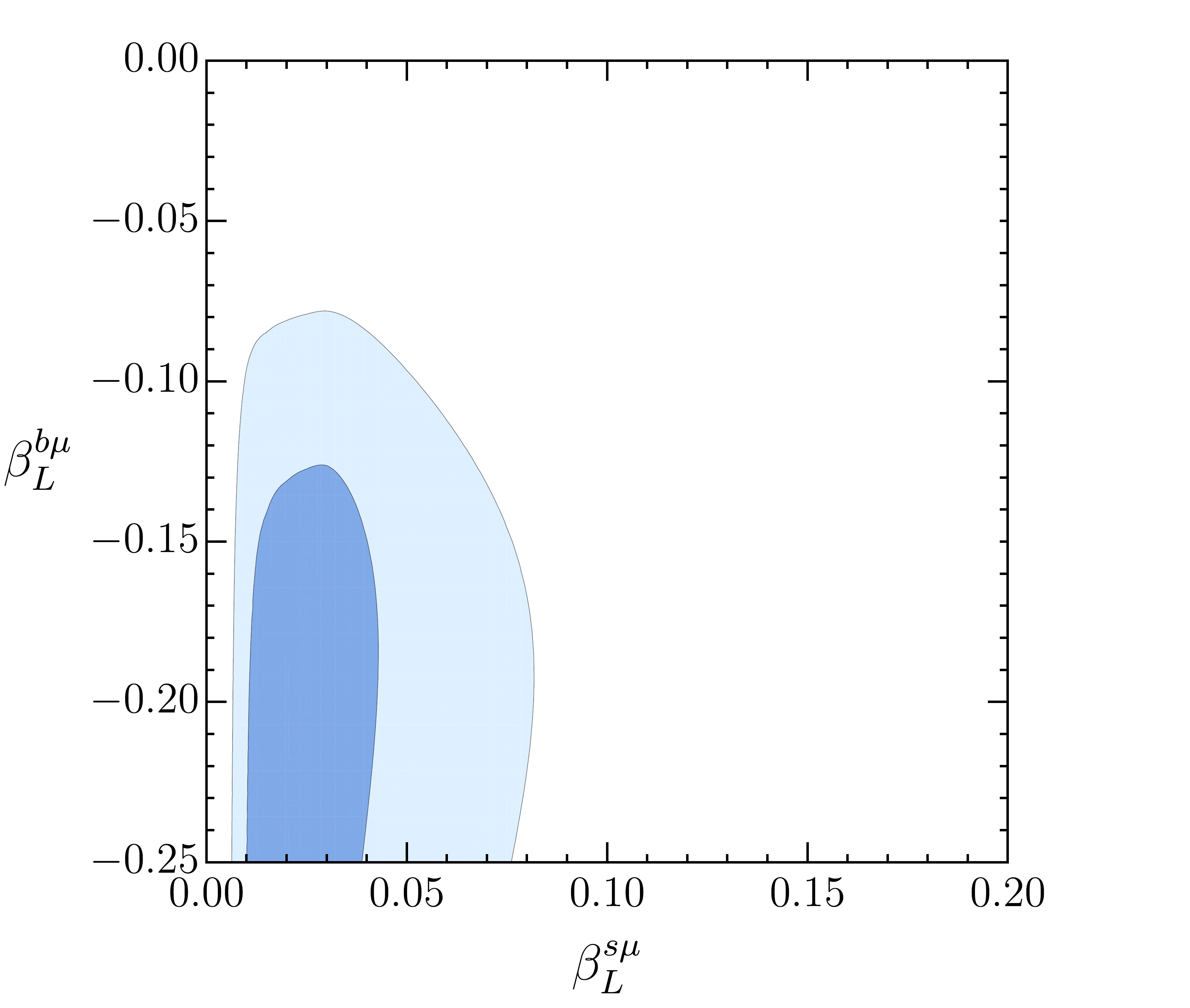}
\caption{Preferred 2D regions from the fit, marginalizing over the rest of the parameters. The $\Delta \chi^{2} \leq 2.30$ ($1\sigma$) and $\Delta \chi^{2} \leq 6.18$ ($2\sigma$) regions are shown in blue and light blue, respectively. Dashed isolines for $\mathcal{B}(B_c\to\tau\nu)$ assuming $R_{D^{(*)}}$  to be fixed to their current experimental central values are also shown. }
\label{fig:2Dfit}
\end{figure}

We perform a fit to low-energy data with five free parameters: $C_{U}$,  $\beta_L^{d\tau}$, $\beta_L^{s \mu}$, $\beta_L^{s \tau}$, and $\beta_L^{b \mu}$.\footnote{For the CKM parameters we use the values from the NP CKM fit from UTfit~\cite{Bona:2017cxr}, and PDG values~\cite{Tanabashi:2018oca} for the rest of the SM parameters. The presence of NP could potentially affect the extraction of these parameters from the experimental observables~(see e.g. \cite{Descotes-Genon:2018foz} for a recent discussion). However, given the flavor structure of our NP (dominantly coupled to third-generation fermions), we do not expect these modifications to significantly alter our fit results, so we neglect these corrections in the following.} The observables entering the fit, together with their SM predictions and experimental values, are given in Table \ref{tab:fitobs}.\footnote{The recent analyses in~\cite{Alguero:2019ptt,Straub:Moriond} indicate a non-vanishing negative value of 
$\mathcal{C}_{9}^{U}$, with different levels of statistical significance. Since $\mathcal{C}_{9}^{U}$ is affected by non-factorizable hadronic 
contributions which are difficult to estimate precisely, we adopt the following conservative choice: 
our 90\% C.L.~upper limit on $\mathcal{C}_{9}^{U}$ is set to 0,  while the 90\% C.L.~lower limit is set to -1.0 (which coincides with the 90\% C.L.~lower limit in~\cite{Alguero:2019ptt}). 
This way the central value of $\mathcal{C}_{9}^{U}$  
is closer to the value quoted in~\cite{Straub:Moriond}, but the error is $\sim 1.5$ times larger.}
 The vector leptoquark contributions to these observables are detailed in Section~\ref{ssec:obs}. We construct the corresponding $\chi^{2}$ and minimize it to obtain the best fit point and best fit regions for the model parameters. Since the observables considered in the fit are not sensitive to the individual signs of $\beta_L^{s\mu}$ and $\beta_L^{b\mu}$ but only to their product (which has to be negative), there is a degeneracy in the fit. We remove this degeneracy by considering $\beta_L^{s\mu}$  to be positive and  $\beta_L^{b\mu}$ negative. We further impose $\beta_L^{s\tau}\leq 0.25$. While the latter condition is not enforced by 
any of the constraints considered here, it finds a natural  justification in the UV-complete model discussed in Section~\ref{sec:UV}. 
As we show in this section, $\beta_L^{s\tau}$ is  the breaking parameter of an approximate flavor symmetry holding at high energy 
and is expected to be small.

We find the following best fit $1\sigma$ regions for the fit parameters (marginalizing over the rest of parameters)
\begin{align}
\begin{aligned}
C_{U} &\in \left[2.8, 6.4\right]\cdot 10^{-3} \,,&&& \beta_L^{s \tau} &\in \left[0.15, 0.25\right] \,,&&&  \beta_L^{d\tau} \in \left[-0.17,-0.02 \right]~, \\
\beta_L^{b \mu} &\in \left[-0.46, -0.16 \right] \,,&&& \beta_L^{s \mu} &\in \left[0.01,0.03 \right] \,.
\end{aligned}
\end{align}
The corresponding 2D $1\sigma$ and $2\sigma$ marginalized contours are shown in Figure~\ref{fig:2Dfit}.  As can be seen, not all the parameters are tightly constrained. However, the $1\sigma$ regions are well compatible with the expected hierarchical structure of the 
$\beta$'s. More precisely, data are compatible with $|\beta_{L}^{s \tau}|,~|\beta_{L}^{b \mu}| = O(10\%)$ and 
 $|\beta_{L}^{d \tau}|,~|\beta_{L}^{s \mu}| = {\rm few}\times 10^{-2}$.

\begin{figure}[t]
\centering
\includegraphics[width=0.42\textwidth]{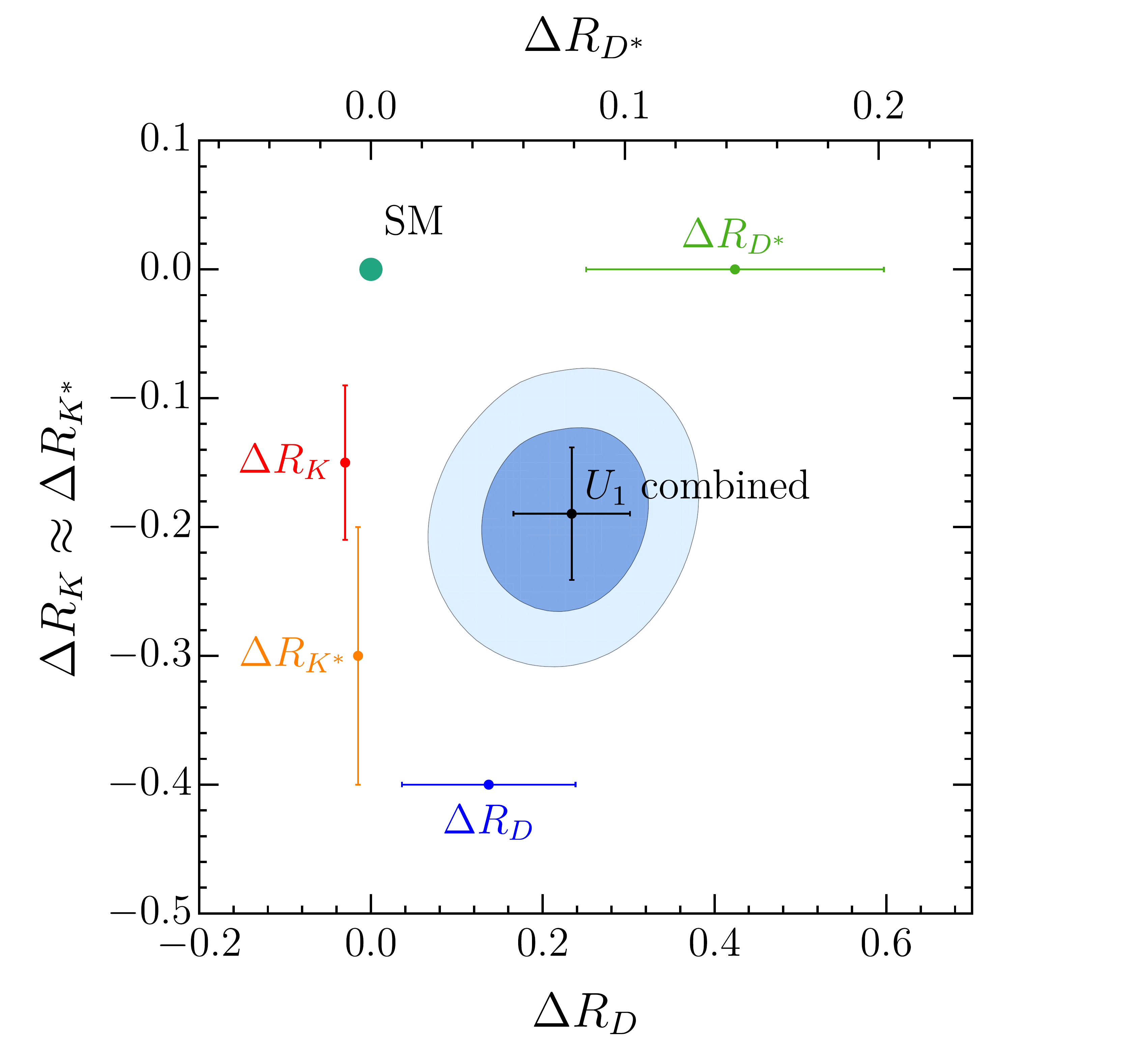} \includegraphics[width=0.42\textwidth]{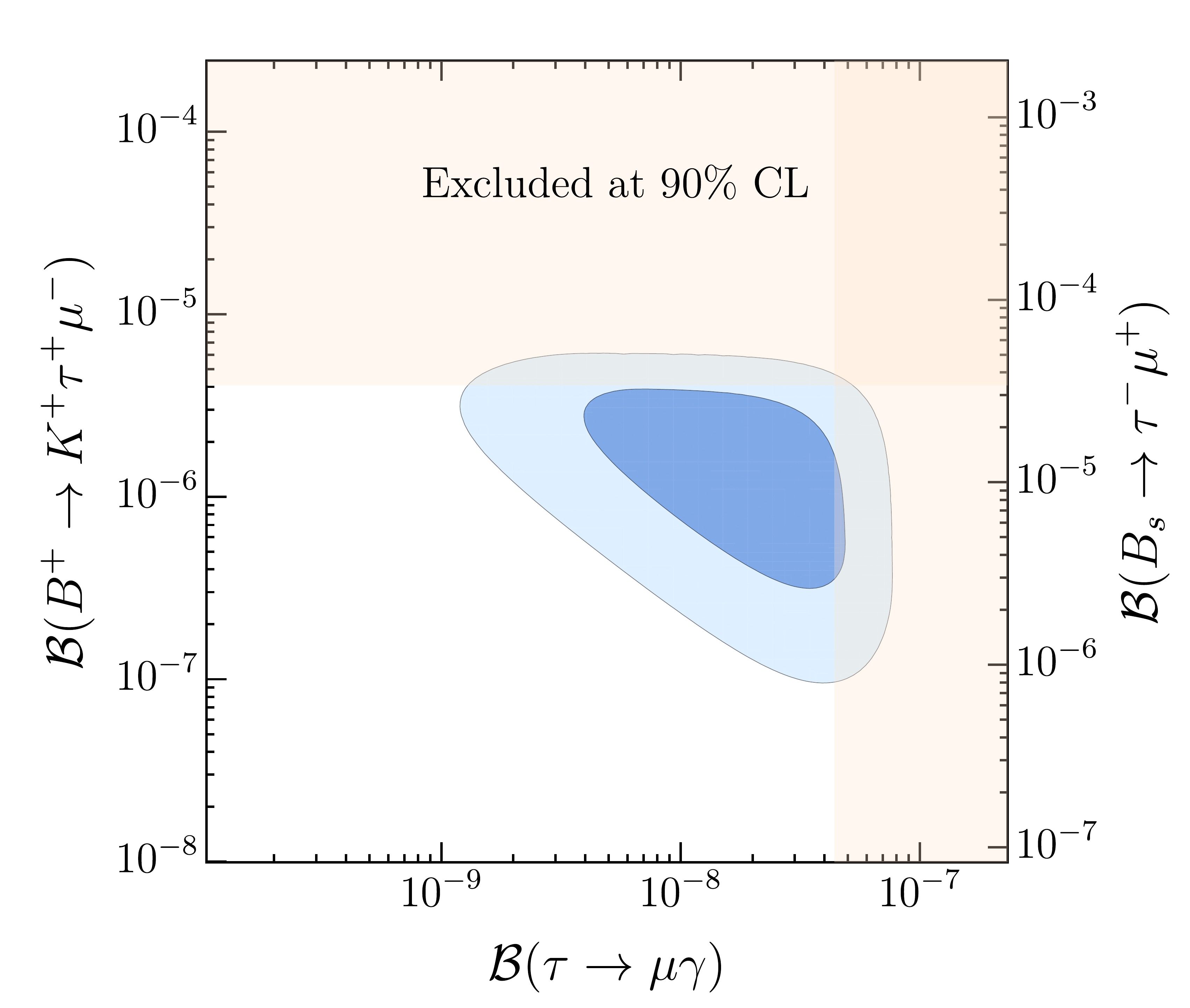} \\ 
\includegraphics[width=0.40 \textwidth]{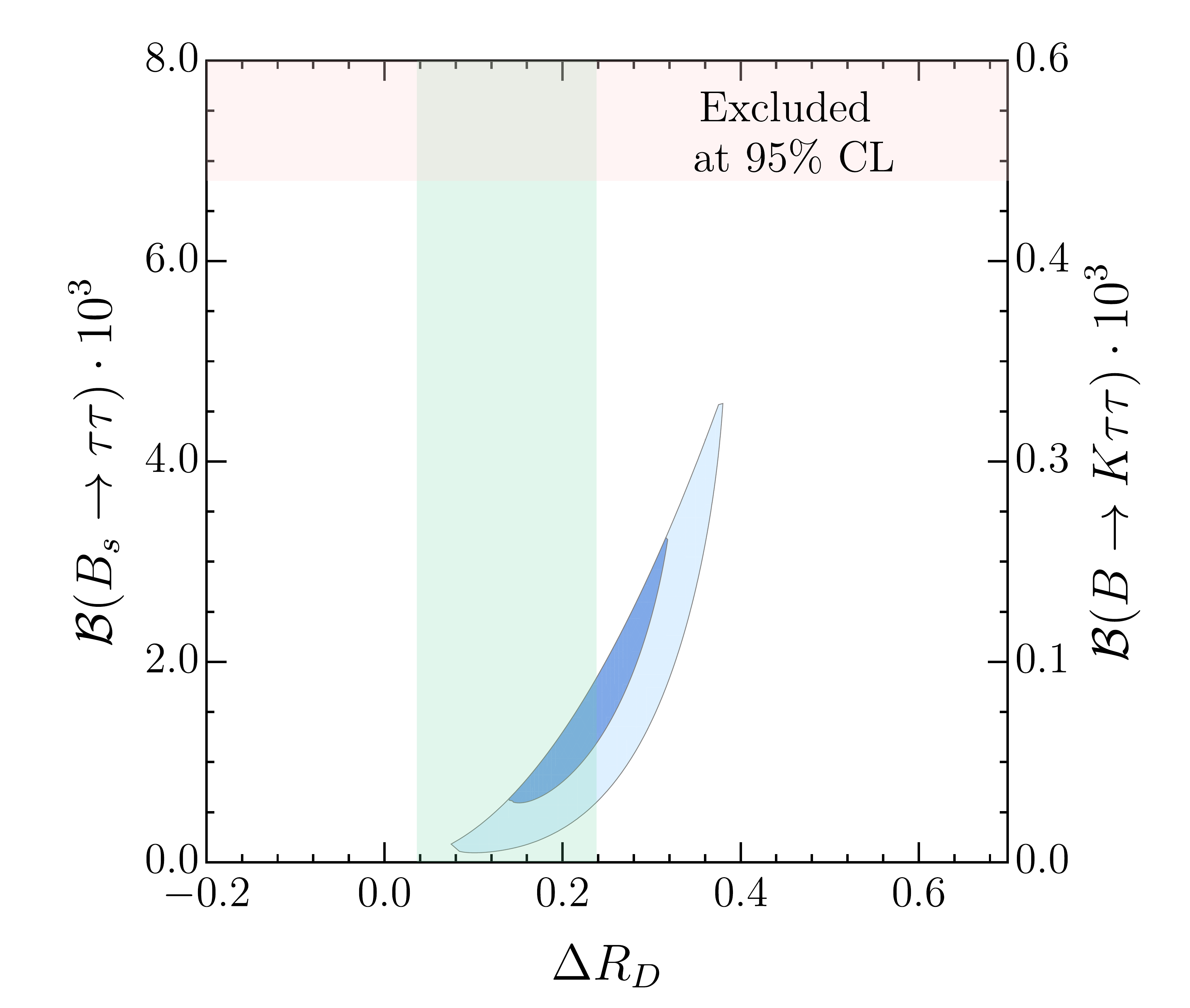}  \includegraphics[width=0.42\textwidth]{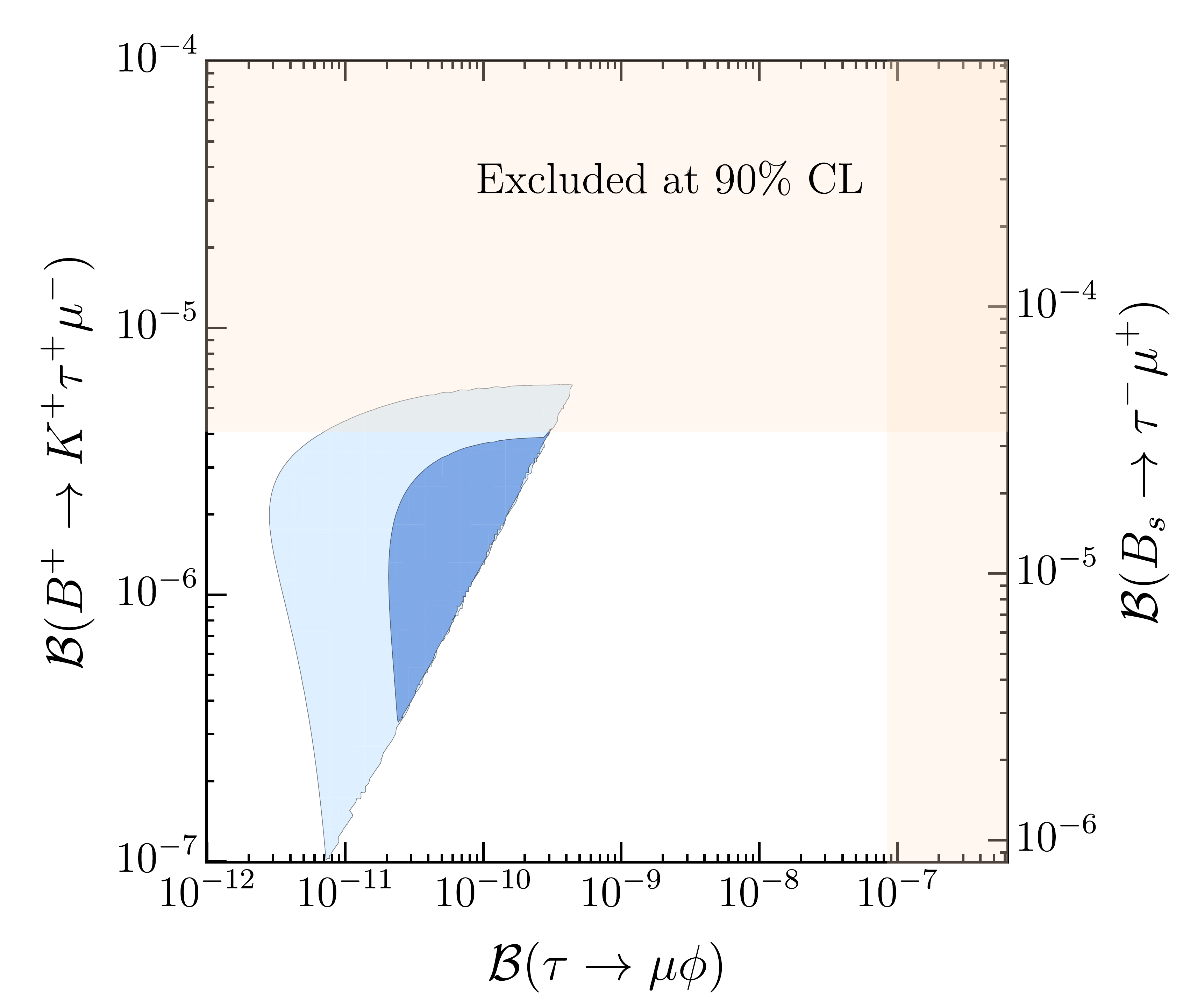}
\caption{Preferred 2D fit regions for different experimental observables, marginalizing over the rest of parameters. The $\Delta \chi^{2} \leq 2.30$ ($1\sigma$) and $\Delta \chi^{2} \leq 6.18$ ($2\sigma$) regions are shown in blue and light blue, respectively. The intervals in the top left plot show the current experimental measurements of $\Delta R_{K^{(*)}}$ and $\Delta R_{D^{(*)}}$ with $1\sigma$ errors. The cross corresponds to the combination of these measurements assuming the relation among these observables in our $U_1$ model. The red (orange) bands show the $95\%\,(90\%)$ CL experimental exclusion limits, while the green band indicates the experimental measurement at $1\sigma$.}
\label{fig:LowE_pheno}
\end{figure}

The main conclusions we can draw from this fit are the following:
\begin{itemize}
\item  $\Delta R_{D^{(*)}}$ fixes the product of $C_{U}$ and $\beta_L^{s\tau}$, and the two are therefore anticorrelated, see Fig. \ref{fig:2Dfit} (top left). The same behavior is also seen in the pure left-handed scenario~\cite{Buttazzo:2017ixm}.  However, in this case the presence of the right-handed coupling yields a significantly larger NP contribution to $\Delta R_{D}$ for fixed $C_U$, allowing for smaller values of $C_{U}$, or equivalently for a larger $M_{U}$ at fixed vector leptoquark coupling $g_U$. The impact of this concerning high-$p_T$ searches is discussed in Section~\ref{ssec:highpT}. As shown in~\cite{Bordone:2018nbg}, the low-region of $\beta_L^{s\tau}$, with correspondingly larger values of $C_U$, receives important constraints from $\tau \to \mu \gamma$, which sets an upper limit in $C_U$ of about $0.02$. On the other hand, we find that the radiative constraints from LFU ratios in $\tau$ decays give comparable limits to those from $\tau\to\mu\gamma$.

\item Given the sizable values of $\beta_L^{s\tau}$ and the chiral enhancement due to $\beta_R^{b\tau}$, we end up with an $\mathcal{O}(10^3)$ NP enhancement in $\mathcal{B}(B_{s} \to \tau^{+} \tau^{-})$ and in $\mathcal{B}(B \to K \tau^{+} \tau^{-})$, within the reach of future experimental limits, see Figure~\ref{fig:LowE_pheno} (bottom left). Improvements in these observables are therefore crucial to test the validity of this setup.

\item A similar scalar enhancement could also yield dangerous NP effects in $\mathcal{B}(B \to \tau \nu)$. Those are however alleviated in the presence of $\beta_L^{d \tau}$, see Fig.~\ref{fig:2Dfit} (bottom left). In particular, while a zero value for $\beta_L^{d \tau}$ is disfavored in our setup, a wide range of non-zero values for $\beta_L^{d \tau}$ are allowed. Interestingly, the relation $\beta_L^{d \tau}=-|V_{td}/V_{ts}|\, \beta_L^{s \tau}$, which is naturally expected in a $U(2)$ framework with a single spurion breaking~\cite{Buttazzo:2017ixm}, is perfectly consistent with our preferred fit region. This relation arises naturally also in the UV completion given in Section~\ref{sec:UV}. 

\item As in the pure left-handed case, the couplings $\beta_L^{b \mu}$ and $\beta_L^{s \mu}$ are anticorrelated and need to be of opposite sign in order to reproduce the measured value of $\Delta C_9^{\mu \mu}=-\Delta C_{10}^{\mu \mu}$, see Fig.~\ref{fig:2Dfit} (bottom right). The maximum size of $\beta_L^{s \mu}$ is mildly constrained by the current experimental bound in $\mathcal{B}(B^{+} \to K^{+} \tau \mu)$. More stringent constraints are obtained by the recent LHCb measurement of $\mathcal{B}(B_s \to \tau \mu)$, for which larger NP effects are expected due to the additional chiral enhancement. 
On the other hand, the maximum size of $\beta_L^{b \mu}$ is bounded by the constraints from $\tau \to \mu \gamma$. Finally, NP contributions to $\tau \to \mu \phi$ are limited by our assumption $\beta_L^{s\tau}\leq0.25$. We find these contributions to be more than two orders of magnitude below the current experimental limits, see Figure~\ref{fig:LowE_pheno} (bottom right).

\item The universal contribution along the $\Delta \mathcal{C}_{9}$ direction is correlated with the NP effect in $R_{D(*)}$. Marginalizing over all other parameters, we find the best fit $1\sigma$ region $\Delta \mathcal{C}_{9}^{U}\in[-0.33, -0.19]$, in reasonable agreement with what is expected from the fit to $b\to s\mu\mu$ observables~\cite{Alguero:2019ptt,Straub:Moriond}.
\end{itemize}

The best fit region in the proposed framework is consistent with a combined explanation of the two LFU anomalies. This is illustrated in Figure~\ref{fig:LowE_pheno} (top left) where we show the $1\sigma$ and $2\sigma$ preferred fit regions for $\Delta R_{K^{(*)}}$ and $\Delta R_{D^{(*)}}$ ($\Delta R_{D^{(*)}}\equiv R_{D^{(*)}}/R_{D^{(*)}}^{\rm SM}-1$),
together with their experimental values. Moreover, our setup predicts interesting implications that connect the NP contributions to the anomalies with other observables. The most remarkable of those involves large (chirally enhanced) NP effects in LFV and in $b\to s\tau\tau$ transitions. As shown in Figure~\ref{fig:LowE_pheno}, the model predictions for several observables concerning these transitions, such as $\tau\to\mu\gamma$, $B\to K\tau\mu$ or $B_s\to\tau\tau$, lie close to the current experimental limits. 

%%%%%%%%%%%%%%%%%%%%%%%%%%%%%%%%%
\subsection{Constraints from high-$p_T$ observables}\label{ssec:highpT}
%%%%%%%%%%%%%%%%%%%%%%%%%%%%%%%%%

Having analyzed the low-energy constraints on the dynamical model introduced in Section~\ref{sec:LQDyn},
 we comment now on the most relevant high-$p_T$ constraints on this setup (with the couplings fixed by the fit presented above).
 To this purpose, we take advantage of the recent analysis in~\cite{Baker:2019sli}, 
 where the high-$p_T$ constraints on a $SU(2)_L$-singlet  
 vector leptoquark have been analyzed in general terms.  
 Similarly to the previous section, we fix $\beta_R^{b\tau}=-1$
 and comment on the main differences with respect to the chiral vector leptoquark solution ($\beta_R^{b\tau}=0$).

\begin{figure}[!t]
\centering
\includegraphics[width=0.45\textwidth]{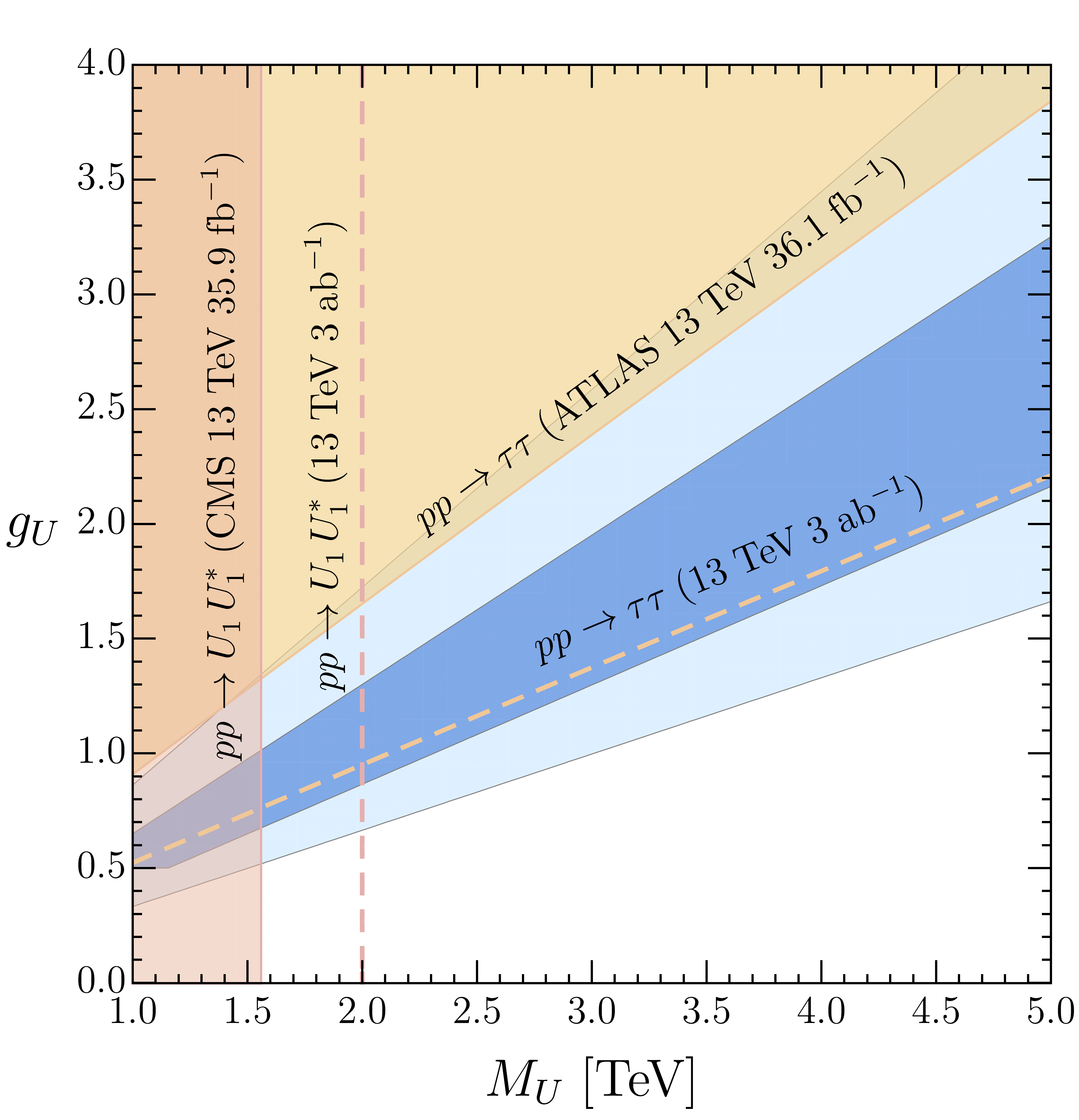} 
\caption{High-$p_T$ constraints on the $SU(2)_L$-singlet vector leptoquark model with $\beta_R^{b\tau}=-1$. The $1\sigma$ and $2\sigma$ regions preferred by the low-energy fit are shown in blue and light blue, respectively.}
\label{fig:highpT_pheno}
\end{figure}

One of the most relevant collider signatures of the model is the production of tau lepton pairs at high energies ($pp\to\tau\tau+X$) via a tree-level t-channel leptoquark exchange. The dominant production mechanism for this channel is through the $b\bar b$ initial state. 
Though slightly pdf enhanced, the production via $b\bar s$ or $s\bar s$ are suppressed by $\beta_L^{s\tau}$.
Due to the smallness of this coupling resulting from the low-energy fit,
 these latter contributions only give a small correction. The most stringent limits in the ditau search are 
 provided by the ATLAS Collaboration with $36.1\,\mathrm{fb}^{-1}$ of $13$~TeV data~\cite{Aaboud:2017sjh}. A recast of the ATLAS search~\cite{Baker:2019sli} shows that a significant region of the parameter space 
 (corresponding to values of $\beta_L^{s\tau}\lesssim0.08 \, (0.03)$ for the $1\sigma$ ($2\sigma$) fit regions) 
 is already excluded by this search, see Figure~\ref{fig:2Dfit}. However, a large portion of the parameter space remains viable. In Figure~\ref{fig:highpT_pheno} we present the current limits, and those obtained 
by extrapolating the statistics to $3\,\mathrm{ab}^{-1}$, assuming that no NP signal will show up and that 
 the SM background uncertainties scale with luminosity as $1/\sqrt{N}$. 
 Interestingly, and in contrast with the chiral vector leptoquark solution (see e.g.~\cite{Angelescu:2018tyl,Schmaltz:2018nls}), 
 the preferred parameter space of the scenario we propose will be almost fully probed by the HL-LHC, provided the current central value for the $R_{D^{(*)}}$ anomaly stays unchanged. 
 This difference is due to the additional contributions from $b_R b_R\to \tau_R\tau_R$ and $b_R b_L\to \tau_R\tau_L$ when $\beta_R^{b\tau}=-1$, which are not fully compensated by the increased NP scale due to the additional scalar contribution in  $R_{D^{(*)}}$.
Analogous limits from $pp\to\tau\mu$ or $pp\to\mu\mu$ are found to be weaker, due to the smallness of $\beta_L^{b\mu}$ and $\beta_L^{s\mu}$, and thus do not play any role in the present discussion~\cite{Baker:2019sli}. Similarly, and in close analogy to what happens in the
chiral leptoquark case~\cite{Greljo:2018tzh}, the corresponding limits from $pp \rightarrow \tau \nu$ are also weaker than the ones from $pp\to\tau\tau$~\cite{Baker:2019sli}. This is due to the smallness of $V_{cs}^*\,\beta_L^{s\tau}$ and $V_{cb}^*\,\beta_L^{b\tau}$, in the present model, compared to $\beta_L^{b\tau}$. 

Complementary constraints can be obtained from bounds on leptoquark pair production, i.e. $pp\to U_1 U_1^*$. Being charged under color, leptoquark pair production is dominated by QCD and therefore it is (almost) independent of the $g_U$ coupling. In our case (with $\beta_R^{b\tau}=-1$), the dominant decay channel of the vector leptoquark is through a $b$-quark and a $\tau$-lepton. The CMS Collaboration has performed a search on $pp\to \tau\tau jj$ with $35.9\,\mathrm{fb}^{-1}$ of data $13$~TeV~\cite{Sirunyan:2018vhk}. Recasting the CMS search one obtains a lower limit in the leptoquark mass of $M_U\gtrsim 1.5$~TeV~\cite{Baker:2019sli}. 
As in the case of $pp\to\tau\tau$ limits, 
in Figure~\ref{fig:highpT_pheno} we report both present and HL-LHC ($3\,\mathrm{ab}^{-1}$) projections 
for the pair-production limits.

%%%%%%%%%%%%%%%%%%%%%%%%%%%%%%%%%%%%%%%%%%%%%%%%%%%%%%%%%%%%%%%%%%
\section{A possible UV completion}\label{sec:UV}
%%%%%%%%%%%%%%%%%%%%%%%%%%%%%%%%%%%%%%%%%%%%%%%%%%%%%%%%%%%%%%%%%%
An important limitation of the phenomenological analysis in Section~\ref{sec:LQDyn} is the inability to reliably estimate
some of the loop contributions that are potentially relevant  for the low-energy phenomenology. Moreover, it is not obvious whether the 
conditions necessary for a successful low-energy fit, and the compatibility with high-$p_T$ constraints,
can be achieved in the context of a consistent UV complete model.
For instance, the non-vanishing value of $\beta_L^{s \tau}$ required by the fit is incompatible with the UV
model proposed in~\cite{Bordone:2017bld}, and even assuming such a large off-diagonal flavor coupling 
can be generated (via a suitable modification of the model), it is not clear if the resulting $B_s$-mixing 
amplitude is in agreement with data. Furthermore, UV-complete models necessarily introduce new particles other than the $U_1$, which could alter the conclusions based on the $U_1$ alone.

We address these questions in this section. To this purpose, we introduce a specific , but sufficiently general,
UV-complete model that allows us to reproduce all the features of the simplified Lagrangian in (\ref{eq:LQLag}).

%%%%%%%%%%%%%%%%%%%%%%%%%%%%%%%%
\subsection{Gauge symmetry and matter content}\label{sec:model}
%%%%%%%%%%%%%%%%%%%%%%%%%%%%%%%%

The model we propose is based on the so-called ``4321"  gauge group, $\mathcal{G}_{4321}\equiv SU(4)\times SU(3)^\prime\times SU(2)_L\times U(1)^\prime$ which contains the SM gauge group as a subgroup.\footnote{As argued in~\cite{Baker:2019sli}, this is the minimal gauge group containing the $U_1$ as a gauge boson which fulfills the necessary requirements to provide a successful explanation of the anomalies while remaining consistent with high-$p_T$ data. See also~\cite{DiLuzio:2017vat} for the first ``4321'' implementation aimed to address the $B$-anomalies, where this point was also noted.} We denote the corresponding gauge fields as $H_\mu^\alpha$, $C_\mu^a$, $W_\mu^i$ and $B^\prime_\mu$, the gauge couplings as $g_4$, $g_3$, $g_L$, $g_1$, and the generators as $T_4^\alpha$, $T_3^a$, $T_L^i$ and $Y^\prime$, with indices $\alpha=1,\dots,15$, $a=1,\dots,8$ and $i=1,2,3$. We normalize the generators so that $\mathrm{Tr}(T^A T^B)=\delta_{AB}$. Many models based in this gauge symmetry have been proposed in the recent literature, see e.g.~\cite{Georgi:2016xhm,Diaz:2017lit,DiLuzio:2017vat,Blanke:2018sro}. In contrast to these proposals, in our model the gauge group is non-universal among the different SM-like families. This (flavored) gauge structure, which can be regarded as a low-energy limit of the PS$^3$ model proposed in~\cite{Bordone:2017bld} (see also~\cite{Bordone:2018nbg,Greljo:2018tuh}), also yields interesting implications in the Yukawa sector of the theory, hinting to a possible explanation of the SM flavor hierarchies. 

\begin{table}[t]
\begin{center}
\begin{tabular}{|c|c|c|c|c|}
\hline
Field & $SU(4)$ & $SU(3)'$ & $SU(2)_L$ & $U(1)'$ \\
\hline
\hline
$q'^i_L$ & $\mathbf{1}$ & $\mathbf{3}$ & $\mathbf{2}$ & $1/6$ \\
$u'^i_R$ & $\mathbf{1}$ & $\mathbf{3}$ & $\mathbf{1}$ & $2/3$  \\
$d'^i_R$ & $\mathbf{1}$ & $\mathbf{3}$ & $\mathbf{1}$ & $-1/3$  \\
$\ell'^i_L$ & $\mathbf{1}$ & $\mathbf{1}$ & $\mathbf{2}$ & $-1/2$ \\
$e'^i_R$ & $\mathbf{1}$ & $\mathbf{1}$ & $\mathbf{1}$ & $-1$ \\ 
$\psi_L^\prime$ & $\mathbf{4}$ & $\mathbf{1}$ & $\mathbf{2}$ & $0$ \\ 
$\psi_u^\prime$ & $\mathbf{4}$ & $\mathbf{1}$ & $\mathbf{1}$ & $1/2$ \\  
$\psi_d^\prime$ & $\mathbf{4}$ & $\mathbf{1}$ & $\mathbf{1}$ & $-1/2$ \\  \rowcolor{RGray}
$\chi^i_L$ & $\mathbf{4}$ & $\mathbf{1}$ & $\mathbf{2}$ & 0  \\  \rowcolor{RGray}
$\chi^i_R$ & $\mathbf{4}$ & $\mathbf{1}$ & $\mathbf{2}$ & 0  \\
\hline
\hline
$H_1$ & $\mathbf{1}$ & $\mathbf{1}$ & $\mathbf{2}$ & 1/2  \\     \rowcolor{RGray}
$H_{15}$ & $\mathbf{15}$ & $\mathbf{1}$ & $\mathbf{2}$ & 1/2  \\  \rowcolor{RGray}
$\Omega_1$ & $\mathbf{\bar 4}$ & $\mathbf{1}$ & $\mathbf{1}$ & $-1/2$  \\ \rowcolor{RGray}
$\Omega_3$ & $\mathbf{\bar 4}$ & $\mathbf{3}$ & $\mathbf{1}$ & $1/6$  \\  \rowcolor{RGray}
$\Omega_{15}$ & $\mathbf{15}$ & $\mathbf{1}$ & $\mathbf{1}$ & 0  \\ 
\hline
\end{tabular}
\end{center}
\caption{ Field content of the model ($i=1,2$). Particles added to the SM matter content are shown on a grey background.   
}
\label{tab:fieldcontent}
\end{table}

The matter content of the theory, together with its representation under $\mathcal{G}_{4321}$, is given in Table~\ref{tab:fieldcontent}. The discussion on the neutrino sector of the theory is beyond the scope of this paper. The observed neutrino masses and mixing angles can be reproduced, without fine-tuning, via an inverse see-saw mechanism by adding additional gauge-singlet fermions~\cite{Greljo:2018tuh} (see also~\cite{Perez:2013osa} for a similar implementation). The fermion content of the model comprises three SM-like and two vector-like families. Two of the SM-like families are singlets under the $SU(4)$ gauge group: $q'^i_L$, $u'^i_R$, $d'^i_R$, $\ell'^i_L$ and $e'^i_R$, with $i=1,2$. The third family SM-like fermions form $SU(4)$ multiplets, $\psi_{L,u,d}$, in which quarks and leptons are unified as $\psi_L^{\prime\intercal}\equiv(q_L^{\prime 3}\;\;\ell_L^{\prime 3})$, $\psi_u^{\prime\intercal}\equiv(u_R^{\prime 3}\;\;\nu_R^{\prime 3})$ and $\psi_d^{\prime\intercal}\equiv(d_R^{\prime 3}\;\;e_L^{\prime 3})$. The vector-like families, $\chi_{L,R}^i\;(i=1,2)$, also form $SU(4)$ multiplets, which decompose under the SM gauge group as $\chi_{L,R}^{i\,\intercal}\equiv(Q_{L,R}^{\prime\,i},L_{L,R}^{\prime\,i})$, where $Q_{L,R}^{\prime\,i}$ and $L_{L,R}^{\prime\,i}$ have the same quantum numbers as the SM $SU(2)_L$ doublets.

The spontaneous symmetry breaking (SSB) of the ``4321'' gauge group down to the SM one is triggered by the vev of $\Omega_{1,3,15}$. While only $\Omega_3$ is enough to trigger the desired symmetry breaking pattern, the additional scalar fields are needed to generate the correct fermion-mixing effects, see Section~\ref{sec:mixing}. An important difference with respect to the models in~\cite{Bordone:2017bld,Bordone:2018nbg,Greljo:2018tuh} is given by the presence of an additional scalar field, $\Omega_{15}$. As we show in the next section, this field plays a key role in generating the 2-3 flavor misalignment in the $U_1$ interactions required by the low-energy fit, see Figure~\ref{fig:2Dfit} (top left). We assume that the scalar potential is such that these scalar fields develop vevs in the following directions
\begin{align}
\langle\Omega^\intercal_{1}\rangle=\frac{1}{\sqrt{2}}
\begin{pmatrix}
0\\
0\\
0\\
\omega_1\\
\end{pmatrix}
\,,\quad
\langle\Omega^\intercal_{3}\rangle&=\frac{1}{\sqrt{2}}
\begin{pmatrix}
\omega_3 & 0 & 0\\
0 & \omega_3 & 0\\
0 & 0 & \omega_3\\
0 & 0 & 0\\
\end{pmatrix}
\,,\quad
\langle\Omega_{15}\rangle=\omega_{15}\,T_4^{15}
\,,
\label{eq:omegadef}
\end{align}
with $\omega_{1,3,15}=\mathcal{O}(\mathrm{TeV})$. These scalar fields can be decomposed under the unbroken SM subgroup as $\Omega_1\sim(\mathbf{\bar 3},\boldsymbol{1})_{-2/3}\oplus (\boldsymbol{1},\boldsymbol{1})_0$, $\Omega_3\sim(\boldsymbol{8},\boldsymbol{1})_0\oplus (\boldsymbol{3},\boldsymbol{1})_{2/3}\oplus (\boldsymbol{1},\boldsymbol{1})_0$, and $\Omega_{15}\sim(\boldsymbol{8},\boldsymbol{1})_0\oplus (\boldsymbol{3},\boldsymbol{1})_{2/3}\oplus (\boldsymbol{1},\boldsymbol{1})_0$. As a result, after removing the Goldstones, we end up with two real color octects, two real and one complex singlets, and two complex leptoquarks. The vector-boson spectrum after SSB, which coincides with the one originally proposed in Ref.~\cite{DiLuzio:2017vat}, contains the following massive fields 
\begin{align}
\begin{aligned}
 & U_\mu^{1,2,3}=\frac{1}{\sqrt{2}}\left(H_{\mu}^{9,11,13}-iH_{\mu}^{10,12,14}\right)~, \quad 
 Z^\prime_\mu = \frac{1}{ \sqrt{g_4^2 + \frac{2}{3}\, g_1^2 }  } \bigg( g_4  H_{\mu}^{15}- \sqrt{\frac{2}{3}} \,g_1  B^\prime_\mu \bigg)~,  \\
&  G_\mu^{\prime\, a} =\frac{1}{ \sqrt{  g_4^2+g_3^2  } } \left( g_4\, H_{\mu}^a - g_3\, C_{\mu}^a \right)~,
\end{aligned}
\end{align}
whose masses read~\cite{DiLuzio:2018zxy}
\begin{align}\label{eq:VectorMasses}
\begin{aligned}
M_{U}&=\frac{1}{2}g_4\sqrt{\omega_1^2+\omega_3^2+\frac{4}{3}\omega_{15}^2}\,, &&& 
M_{Z^\prime}&=\frac{1}{2}\sqrt{\frac{3}{2} g_4^2+g_1^2}\sqrt{\omega_1^2+\frac{1}{2}\omega_3^2}\,, &&&
M_{G^\prime}&=\sqrt{\frac{g_4^2+g_3^2}{2}}\,\omega_3\,.
\end{aligned}
\end{align}
The orthogonal combinations to $G_\mu^{\prime\, a}$ and $ Z^\prime_\mu$ correspond to the SM gauge fields $G_\mu^a$ and $ B_\mu$, whose couplings are $g_c=g_3 g_4/\sqrt{  g_4^2+g_3^2 }  $ and $g_Y=  g_1 g_4 / \sqrt{g_4^2 + \frac{2}{3}\, g_1^2 }$. In particular, the SM color group corresponds to $SU(3)_c\equiv[SU(3)_4\times SU(3)^\prime]_{\rm diag}$ and $U(1)_Y\equiv[U(1)_4\times U(1)^\prime]_{\rm diag}$, with $SU(3)_4\times U(1)_4\subset SU(4)$. In turn, hypercharge is given in terms of the original gauge generators and $U(1)^\prime$-charges by $Y=\sqrt{2/3}\,T^{15}_4+Y^\prime$. The $SU(2)$ group remains unaffected and directly corresponds to the SM $SU(2)_L$.

The two remaining scalar fields, $H_{1,15}$, are responsible of electroweak symmetry breaking (EWSB). The $H_{15}$ field decomposes under the SM gauge group as $H_{15}\sim(\boldsymbol{8},\boldsymbol{2})_{1/2}\oplus (\boldsymbol{3},\boldsymbol{2})_{7/6}\oplus (\mathbf{\bar 3},\boldsymbol{2})_{-1/6}\oplus (\boldsymbol{1},\boldsymbol{2})_{1/2}$, and therefore contains a Higgs doublet which we denote by $H_{15}^0$. This additional Higgs field is needed to generate a (small) splitting between the quark and lepton masses of the would-be third family.  We assume that the scalar potential is such that only $H_{1}$ and $H_{15}^0$ acquire a vev around the electroweak scale. Namely, $|\langle H_1\rangle|=v_1/\sqrt{2}$ and $|\langle H_{15}^0\rangle|=v_{15}/\sqrt{2}$, with the SM vev being given by $v=\sqrt{v_1^2+v_{15}^2}$.  

The leptoquarks in the model do not mediate proton decay at the renormalizable level since they do not couple to quark pairs. As in the SM, baryon and lepton number arise as accidental global symmetries and proton decay can only happen at the level of dimension-six (or higher) operators.

%%%%%%%%%%%%%%%%%%%%%%%%%%%%%%%%
\subsection{Flavor symmetries and fermion-mixing structure}\label{sec:mixing}
%%%%%%%%%%%%%%%%%%%%%%%%%%%%%%%%

In the absence of vector-like fermions, only $\psi_{L,u,d}^\prime$, which we identify with the would-be third family, couple to the $U_1$. As required by gauge anomaly cancellation, both left- and right-handed fermions need to be charged under $SU(4)$, thus the $U_1$ couples to both fermion chiralities with the same coupling strength. The other two SM-like families, being $SU(4)$ singlets, couple to the $Z^\prime$ and $G^\prime$ but not to the $U_1$. We therefore have
\begin{align}\label{eq:ULag}
\mathcal{L}_U&\supset\frac{g_4}{\sqrt{2}}\, U^{\mu}_{1}  \left[ \beta_L^\prime\,(\bar \Psi_q^\prime \gamma_{\mu} P_L \Psi_{\ell}^\prime) +  \beta_u^\prime\,(\bar u_R^\prime \gamma_{\mu} \nu_R^\prime) + \beta_d^\prime\,(\bar d_R^\prime \gamma_{\mu} e_R^\prime) + (\bar Q_R^{\prime\,i} \gamma_{\mu} L_R^{\prime\,i})  \right] + \mathrm{h.c.}\,,
\end{align}
where $\Psi_q^\intercal=(q_L^{\prime\,1}\;q_L^{\prime\,2}\;q_L^{\prime\,3}\; Q_L^{\prime\,1}\; Q_L^{\prime\,2})$, $\Psi_\ell^\intercal=(\ell_L^{\prime\,1}\;\ell_L^{\prime\,2}\; \ell_L^{\prime\,3}\;  L_L^{\prime\,1}\;  L_L^{\prime\,2})$, $\beta_L^\prime=\mathrm{diag}(0,0,1,1,1)$, $\beta_{u,d}^\prime=\mathrm{diag}(0,0,1)$. This is a good starting point to reproduce the solution found in Section~\ref{sec:LQDyn}. Sub-leading couplings to the light generations can then be induced via mass-mixing with the two vector-like families. Given our choice of quantum numbers for the vector-like fermions, mixing effects can only appear in the left-handed sector (before EWSB).  In what follows we discuss these effects, paying special attention to the flavor symmetries of the model. 

In the absence of Yukawa interactions, the fermion sector of the model has the accidental flavor symmetry
\begin{align}\label{eq:flavSym}
\mathcal{G}_F\equiv U(2)_q\times U(2)_{u_R}\times U(2)_{d_R}\times U(2)_\ell\times U(2)_{e_R}
\times U(1)_{\psi_u}\times U(1)_{\psi_d} \times U(3)_{\psi_L+\chi_L}\times  U(2)_{\chi_R}    \,.
\end{align}
We assume that $U(3)_{\psi_L+\chi_L}\times U(2)_{\chi_R}$ is explicitly broken to $U(2)_{\chi}\times U(1)_{\psi_L}$, where
$U(2)_{\chi}\equiv U(2)_{\chi_L+\chi_R}$, by the vector-like mass term. In other words, we assume that the vector-like mass term for $\chi$  is proportional to the identity matrix. While departures from this assumption are possible, $U(2)_{\chi}$-breaking terms are severely constrained, in our model, by $\Delta F=2$ observables. We therefore stick to this assumption for simplicity and consider possible $U(2)_{\chi}$-breaking terms as small perturbations around this solution. The remaining flavor symmetry is explicitly broken by the (renormalizable) Yukawa interactions. Let us analyze these interactions separately. 

We focus first on the Yukawa terms involving the $\Omega_{1,3}$ fields. Without loss of generality, we can use the remaining flavor symmetry to rotate to a basis in which these interactions take the form
\begin{align}\label{eq:O13Chi}
\begin{aligned}
-\mathcal{L} & \supset   M_\chi\,\bar\chi_L\,\chi_R   +  \hat \lambda_{q} \, \bar q_{L} \Omega_{3} \chi_{R}  + \hat \lambda_{\ell} \, W \, \bar \ell_{L} \Omega_{1} \chi_{R}  + \textrm{h.c.} \,,
\end{aligned}
\end{align}
where $M_\chi$ is a (flavor-universal) mass term and $W$, $\hat \lambda_{q}$ and $\hat \lambda_{\ell}$ are $2\times 2$ matrices in flavor space, with $\hat \lambda_{q,\ell}$ diagonal and $W$ unitary. After SSB, these terms induce a mass-mixing between the vector-like and the first- and second-generation SM-like fermions. This way we effectively introduce (small) couplings between the new vectors and the light-generation fermions. We parametrize $\hat \lambda_{q,\ell}$ as follows
\begin{align}\label{eq:lambda_hat}
\begin{aligned}
\hat \lambda_q&=\mathrm{diag}(\lambda_q+\delta\lambda_q,\,\lambda_q)\,,\\
\hat \lambda_\ell&=\mathrm{diag}(\delta\lambda_\ell,\,\lambda_\ell)\,.
\end{aligned}
\end{align}
If considered separately, the parameters $\lambda_{q,\ell}$ yield the following explicit breaking of the flavor symmetry
\begin{align}
\begin{aligned}
U(2)_q\times U(2)_{\chi} &\stackrel{\lambda_q}{\xrightarrow{\hspace*{0.5cm}}} U(2)_{q+\chi}\,,\\
U(2)_\ell\times U(2)_{\chi} &\stackrel{\lambda_\ell}{\xrightarrow{\hspace*{0.5cm}}} U(1)_{\ell_1}\times U(1)_{\chi_1} \times U(1)_{\ell_2+\chi_2}\,.
\end{aligned}
\end{align}
The parameter $\delta\lambda_q$ denotes a possible sub-leading term (i.e. $\delta\lambda_q\ll \lambda_q$) that introduces a small explicit breaking of the $U(2)_{q+\chi}$ symmetry, and is tightly constrained by $\Delta F=2$ observables. Similarly, $\delta\lambda_\ell$  corresponds to a possible sub-leading term (i.e. $\delta\lambda_\ell\ll \lambda_\ell$) that explicitly breaks the $U(1)_{\ell_1}\times U(1)_{\chi_1}$ symmetry, and is constrained by LFV observables such as $K_L\to\mu e$. The simultaneous presence of $\lambda_q$ and $\lambda_\ell$ yields the collective breaking of $U(2)_q\times U(2)_\ell\times U(2)_\chi$. However, since both couplings are required for this breaking to take place, the full breaking of the flavor symmetry (and in particular the flavor misalignment parametrized by $W$) is only seen in the $U_1$ interactions, while the $Z^\prime$ and $G^\prime$ couplings still respect (at tree-level) the $U(2)_q$ and $U(1)_{\ell_1}\times  U(1)_{\chi_1} \times U(1)_{\ell_2+\chi_2}$ symmetries separately. This is analogous to the SM case with the CKM matrix and corresponds to the ``Cabbibo mechanism" described in~\cite{DiLuzio:2018zxy}. However, in contrast to the setup in~\cite{DiLuzio:2018zxy}, in our case this mechanism does not let us induce non-diagonal $U_1$ couplings among second- and third-generation SM fermions, but only among the light-families. For simplicity in the discussion, and in order to avoid large NP contributions to $\Delta F=2$ observables and LFV transitions involving electrons, we set $W=\mathbb{1}$ and $\delta\lambda_q=\delta\lambda_\ell=0$, enhancing the surviving flavor symmetry to $U(1)_{\ell_1}\times U(1)_{q_1+\chi_1}$. As a result, after SSB we obtain the following $U_1$ couplings in the fermion mass-eigenbasis
\begin{align}\label{eq:betaO13}
\begin{aligned}
\beta_{L}^\prime\stackrel{\langle \Omega_{1,3}\rangle}{\xrightarrow{\hspace*{0.8cm}}}\beta_{L} &= \mathcal{R}_{14}(\theta_{q_1})\, \mathcal{R}_{25}(\theta_{q_2}) \,\mathrm{diag}(0,0,1,1,1)\, \mathcal{R}_{25}^\dagger(\theta_{\ell_2}) \\[5pt]
&=
\left(\begin{array}{ccc:cc}
 0 & 0 & 0 & -s_{q_1} & 0 \\
 0 &  s_{\ell_2} s_{q_2} & 0 & 0 & -c_{\ell_2} s_{q_2} \\
 0 & 0 & 1 & 0 & 0  \\[2pt] \hdashline 
 0 & 0 & 0 & c_{q_1} & 0 \\
 0 & -s_{\ell_2} c_{q_2} &0 & 0 & c_{\ell_2} c_{q_2}  \\
\end{array}\right) 
\,,
\end{aligned}
\end{align}
where $\mathcal{R}_{ij}(\theta)$ denotes a rotation of angle $\theta$ between the fermions $i$ and $j$, and $s_{q_i,\ell_2}$ and $c_{q_i,\ell_2}$ are short for $\sin\theta_{q_i,\ell_2}$ and $\cos\theta_{q_i,\ell_2}$.  
The dashed lines in the matrix separate the $3\times 3$ subsector of the chiral 
(SM) fermions from the vector-like ones.
The mixing angles are defined in terms of Lagrangian parameters as
\begin{align}\label{eq:qlmixNo15}
\begin{aligned}
\tan\theta_{q_1}&=\tan\theta_{q_2}=\frac{\lambda_q\,\omega_3}{\sqrt{2} M_\chi}\,,&\quad&&\tan\theta_{\ell_2}&=\frac{\lambda_\ell\,\omega_1}{\sqrt{2}M_\chi}\,.
\end{aligned}
\end{align}
The coupling structure in~\eqref{eq:betaO13} coincides with the one obtained in~\cite{Bordone:2017bld,Bordone:2018nbg,Greljo:2018tuh} before EWSB. As argued in Section~\ref{ssec:fit}, this coupling structure is not enough to provide a good fit to data since a sizable $\beta_L^{23}$ coupling is required. The $2$-$3$ misalignment can be achieved with the Yukawa interactions involving $\Omega_{15}$. We have
\begin{align}\label{eq:O15Chi}
\begin{aligned}
-\mathcal{L} & \supset  \hat \lambda_{15}\, \bar \psi_{L}^\prime \Omega_{15} \chi_{R} + \hat \lambda_{15}^{\prime}\, \bar \chi_{L} \Omega_{15} \chi_{R}  + \textrm{h.c.} \,,
\end{aligned}
\end{align}
where $\hat \lambda_{15}^{\prime}$ is a $2\times 2$ matrix and $\hat \lambda_{15}$  a 2-dimensional vector that we assume to be aligned with the second family, namely  $\hat\lambda_{15}^\intercal=( 0\;\lambda_{15})$.\footnote{Other orientations of $\hat\lambda_{15}$ can be reabsorbed into a redefinition of $\chi_{L,R}$ and do not affect the interactions discussed here.} As  we did with the vector-like mass, we further assume $\hat \lambda_{15}^{\prime}$ to be flavor universal, i.e. we fix $\hat \lambda_{15}^{\prime}=\lambda_{15}^{\prime}\,\mathbb{1}_{2\times 2}$. After SSB, the Lagrangian term proportional to $\hat \lambda_{15}^{\prime}$ generates a mass splitting between vector-like quarks and leptons,
\begin{align}\label{eq:Mchisplit}
M_Q&=M_\chi+\frac{1}{2\sqrt{6}}\,\lambda_{15}^\prime\,\omega_{15}\,,&&&M_L&=M_\chi-\frac{3}{2\sqrt{6}}\,\lambda_{15}^\prime\,\omega_{15}\,.
\end{align}
On the other hand, the parameter $\lambda_{15}$ acts as a new source of flavor breaking. After $\Omega_{15}$ takes a vev, it triggers the following explicit flavor symmetry breaking
\begin{align}
U(2)_{\chi} \stackrel{\lambda_{15}}{\xrightarrow{\hspace*{0.5cm}}} U(1)_{\chi_1}\,.
\end{align}
Analogously to the case discussed above, after SSB the term proportional to $\lambda_{15}$ yields a mass-mixing between the third-generation and one of the vector-like fermions. However, since $T_4^{15}$ commutes with the generators associated to the $Z^\prime$ and $G^\prime$, this breaking is only seen by the $U_1$ interactions, up to very small corrections. This way we are able to generate large non-diagonal $U_1$ interactions between $\psi_{L}^{3}$ and $\chi^2$, proportional to $\lambda_{15}$, while in first approximation (i.e. to first order in the 
flavor-breaking terms)  the $Z^\prime$ and $G^\prime$ interactions remain unaffected, see Appendix~\ref{app:VecCoup}. In combination with the mixing induced by the $\lambda_{q,\ell}$ terms, this translates into sizable contributions to $\beta_L^{23,32}$, while keeping flavor-changing neutral currents under control. More precisely, after SSB we end up with the following $U_1$ interactions in the fermion mass basis
\begin{align}\label{eq:betaO1315}
\begin{aligned}
\beta_{L}^\prime\stackrel{\langle \Omega_{1,3,15}\rangle}{\xrightarrow{\hspace*{1.2cm}}}\beta_{L}&\approx \mathcal{R}_{14}(\theta_{q_1}) \, \mathcal{R}_{25}(\theta_{q_2})\,\mathcal{R}_{35}(\chi_q) \,\mathrm{diag}(0,0,1,1,1)\,\mathcal{R}_{35}^\dagger(\chi_\ell)\,  \mathcal{R}_{25}^\dagger(\theta_{\ell_2}) \\[5pt]
&=
\left(\begin{array}{ccc:cc}
 0 & 0 & 0 & -s_{q_1} & 0 \\
 0 &  s_{\ell_2} s_{q_2}  c_{\chi} & s_{q_2} s_{\chi} & 0 & -c_{\ell_2} s_{q_2} c_{\chi} \\
 0 & -s_{\ell_2} s_{\chi} & c_{\chi} & 0 & c_{\ell_2} s_{\chi}    \\[2pt] \hdashline 
 0  & 0 & 0 & c_{q_1} & 0  \\  
 0 & -s_{\ell_2} c_{q_2} c_{\chi} & -c_{q_2} s_{\chi} & 0 & c_{\ell_2} c_{q_2} c_{\chi} \\
\end{array}\right) 
\,, 
\end{aligned}
\end{align}
where $\chi\equiv\chi_\ell-\chi_q$ and the new mixing angles are related to the Lagrangian parameters by
\begin{align}\label{eq:chiMix}
\tan\chi_q=\frac{1}{2\sqrt{6}}\frac{\lambda_{15}\,\omega_{15}}{M_Q}\,,\qquad\qquad\tan\chi_\ell=\frac{-3}{2\sqrt{6}}\frac{\lambda_{15}\,\omega_{15}}{M_L}\,.
\end{align}
Note that at this point the expressions for $\theta_{q_i,\ell_i}$ are slightly modified compared to those in~\eqref{eq:qlmixNo15}.
The new expressions read
\begin{align}\label{qlmix}
\begin{aligned}
\tan\theta_{q_1}&=\frac{\lambda_q\,\omega_3}{\sqrt{2} M_Q}\,,&\quad&&\tan\theta_{q_2}&=\frac{\lambda_q\,\omega_3}{\sqrt{2} M_Q}\,c_{\chi_q}\,,&\quad&&\tan\theta_{\ell_2}&=\frac{\lambda_\ell\,\omega_1}{\sqrt{2}M_L}\,c_{\chi_\ell}\,.
\end{aligned}
\end{align}
Finally, the resulting physical masses for the vector-like fermions are given by
\begin{align}
\begin{aligned}
M_{Q_1}&=\sqrt{M_Q^2+\frac{|\lambda_q|^2\,\omega_3^2}{2}}\,, &\quad&& M_{Q_2}&=\sqrt{M_Q^2+\frac{|\lambda_q|^2\,\omega_3^2}{2}+\frac{|\lambda_{15}|^2\,\omega_{15}^2}{24}}\,,\\[5pt]
M_{L_1}&=M_L\,,&&&M_{L_2}&=\sqrt{M_L^2+\frac{|\lambda_\ell|^2\,\omega_1^2}{2}+\frac{3\,|\lambda_{15}|^2\,\omega_{15}^2}{8}}\,.
\end{aligned}
\end{align}

After EWSB, a final rotation to bring the SM fermions to their mass-eigenbasis is needed. The Yukawa interactions involving the Higgsses introduce new sources of breaking of the flavor symmetry in~\eqref{eq:flavSym}, whose structure fits well with the minimal $U(2)$ picture in~\cite{Barbieri:2011ci}. A detailed discussion of these symmetry-breaking terms and their connection to the SM fermion masses and mixing angles can be found in~\cite{Bordone:2018nbg} (see also~\cite{Greljo:2018tuh}). In particular, the rotation matrices that bring the SM fermions from the flavor basis defined in~\eqref{eq:DownBasis} to the mass-eigenbasis, $L_{d,\ell}$ and $R_{u,d,e}$, can be found in the Appendix A of~\cite{Bordone:2018nbg}. In this reference, the different breaking of the flavor symmetry are parametrized in terms of new mixing angles whose phenomenological constraints are also discussed. Here, for simplicity, we take $s_b=s_e=\phi_\tau=0$ and fix $\alpha_d=\pi$ in these rotation matrices.\footnote{As shown in~\cite{Bordone:2018nbg}, (small) deviations from these values are possible and might even be welcome in the case of $s_b$ if we allow for the CP violating phase $\phi_b\approx\pi/2$.} Under these assumptions, the $U_1$ interactions in the mass basis for SM fermions can finally be written as
\begin{align}\label{eq:U1coup}
\beta_{L}^\prime\stackrel{\langle \Omega_{1,3,15}\rangle\,,\;\langle H_{1,15}\rangle}{\xrightarrow{\hspace*{2.4cm}}}\beta_{L}&\approx
{
\footnotesize
\left(\begin{array}{ccc:cc}
 0 & -|V_{td}/V_{ts}|\;  c_d\, s_{\ell_2} s_{q_2}  c_{\chi} & -|V_{td}/V_{ts}|\; c_d\, s_{q_2} s_{\chi} & -c_{d}\, s_{q_1} &  -|V_{td}/V_{ts}|\, c_{d}\, c_{\ell_{2}} s_{q_{2}} c_{\chi}  \\
 0 &  c_d \, s_{\ell_2} s_{q_2}  c_{\chi} & c_d \, s_{q_2} s_{\chi}  &   -|V_{td}/V_{ts}|\, c_{d}\, s_{q_{1}} &     - c_{d}\, c_{\ell_2} s_{q_2} c_{\chi}\\
 0 & -s_{\ell_2} s_{\chi}- s_\tau \, c_\chi & c_{\chi} & 0 & c_{\ell_2} s_{\chi} \\[2pt] \hdashline 
 0 & 0 & 0 & c_{q_1} & 0 \\
 0 &     -c_{q_{2}} (s_{\ell_{2}} c_{\chi}  - s_{\tau}\, s_{\chi} )  & -c_{q_{2}}  s_{\chi}  & 0 & c_{\ell_2} c_{q_2} c_{\chi} \\
\end{array}\right) 
}
\,, \nonumber\\[5pt]
\beta_d^\prime\stackrel{\langle \Omega_{1,3,15}\rangle\,,\;\langle H_{1,15}\rangle}{\xrightarrow{\hspace*{2.4cm}}}\beta_d&\approx
e^{i\phi_d}
\left(\begin{array}{ccc}
 0 & 0 & 0  \\
 0 &  0 & 0 \\
 0 & \frac{m_\mu}{m_\tau}\, s_\tau & 1
\end{array}\right) 
\,,
\end{align}
where $c_d\approx0.98$ and $\phi_d$ is an arbitrary phase that we fix to $\phi_d=\pi$ to maximize the NP contributions to $R(D^{(*)})$ (see~\eqref{eq:RD} and~\eqref{eq:RDs}). We stress that the latter choice does need to be enforced and, in presence 
of a more precise measurement of $\Delta R_D/\Delta R_{D^*}$ and/or polarization observables in $b\to c\tau\nu$ transitions, the value of $\phi_d$ could also be extracted from the low-energy fit. 

This flavor structure for the $U_1$ couplings nicely matches the one discussed in Section~\ref{sec:LQDyn} (with $\beta_R\equiv\beta_d$). The only difference between the two structures is given by the non-zero values for  $\beta_L^{d\mu} $ and $\beta_R^{b\mu}$, which were set to zero in \eqref{eq:minCoup}. These two couplings are extremely suppressed (or can be chosen to be very small), justifying a posteriori having neglected them in Section~\ref{sec:LQDyn}. In particular one has $|\beta_L^{d\mu}| = |V_{td}/V_{ts}| |\beta_L^{s\mu}| \lsim  0.01$ (taking into account the fit result for $|\beta_L^{s\mu}|$).\footnote{Such a value of  
$\beta_L^{d\mu}$ has no impact on the low-energy observables considered so far. It would have an impact in $b\to d \ell\ell$ transitions,
if these were measured more precisely in the future: there we expect corrections relative to the SM of the same order as in 
$b\to s \ell\ell$, given the $U(2)_q$ relation $|\beta_L^{d\mu}/\beta_L^{s\mu}| = |V_{td}/V_{ts}|$.
Similar effects in short-distance $s\to d \ell\ell$ amplitudes (contributing e.g.~to 
$K_L\to \mu\mu$) are obscured by long-distance contributions and are, in practice, not detectable.}
The size of $\beta_R^{b\mu}$ is not precisely fixed,  but the phenomenological condition $|\beta_R^{b\mu}/\beta_L^{b\mu} | \lsim 0.02$ (see Section~\ref{ssec:obs}) can be obtained by imposing $|s_\tau |\lsim 0.05$.

At this point we can address more precisely the question of which are the ingredients necessary to generate a sufficiently large 
$\beta^{s\tau}_{L}$. For the purpose of illustration, working in the limit of small mixing angle (i.e. small~flavor symmetry-breaking terms), we get 
\begin{equation}\label{eq:bst_approx}
\beta^{s\tau}_{L} \approx  (\chi_\ell - \chi_q)  \theta_{q_2}  =  -   \lambda_{15} \lambda_{q}  
\frac{  \omega_{3}\, \omega_{15}  }{\sqrt{3}\, M_L\, M_Q} \left[ 1 + \mathcal{O}( \lambda_{15}^\prime) \right]\,.
\end{equation}
As expected, $\beta^{s\tau}_{L}$ is proportional to the two flavor breaking parameters $\lambda_{15}$ and 
$\lambda_{q}$, whose collective presence leads to the effective breaking 
of the $U(2)_q$ symmetry in the $U_1$ couplings.  As we show in the next section, the maximal size of 
these breaking terms is constrained by $\Delta F=2$ amplitudes.

%%%%%%%%%%%%%%%%%%%%%%%%%%%%%%%%
\subsection{Vector leptoquark loops in the UV-complete model}\label{sec:loops}
%%%%%%%%%%%%%%%%%%%%%%%%%%%%%%%%

We are now ready to compute the relevant one-loop effects mediated by the vector leptoquark. An interesting property of the $U_1$ couplings obtained in~\eqref{eq:U1coup}, arising as a consequence of the unitarity of the fermion-mixing matrices, is that $(\beta_L^\dagger \beta_L)_{ij}$ and $(\beta_L \beta_L^\dagger)_{ij}$ are diagonal in the SM sub-block, i.e. for $i,j=1,2,3$. This property, analogous to the GIM mechanism in the SM, ensures a ``flavor protection'' in the $U_1$ loops. Thanks to this protection, we find that $U_1$ contributions to purely leptonic processes such as $\tau\to 3\mu$ and $\tau \to \mu \nu \nu$, or to semileptonic processes like $B\to K^{(*)}\nu\nu$, do not have a relevant phenomenological impact (see also~\cite{DiLuzio:2018zxy,Crivellin:2018yvo,Buttazzo:2017ixm}) and hence we do not include them in our discussion. Instead, we focus here on $\Delta F=2$ and dipole transitions, which are more severely constrained.

\subsubsection{$\Delta F=2$ transitions}\label{sec:DF2}

\begin{figure}[!t]
\centering
\includegraphics[width=0.45\textwidth]{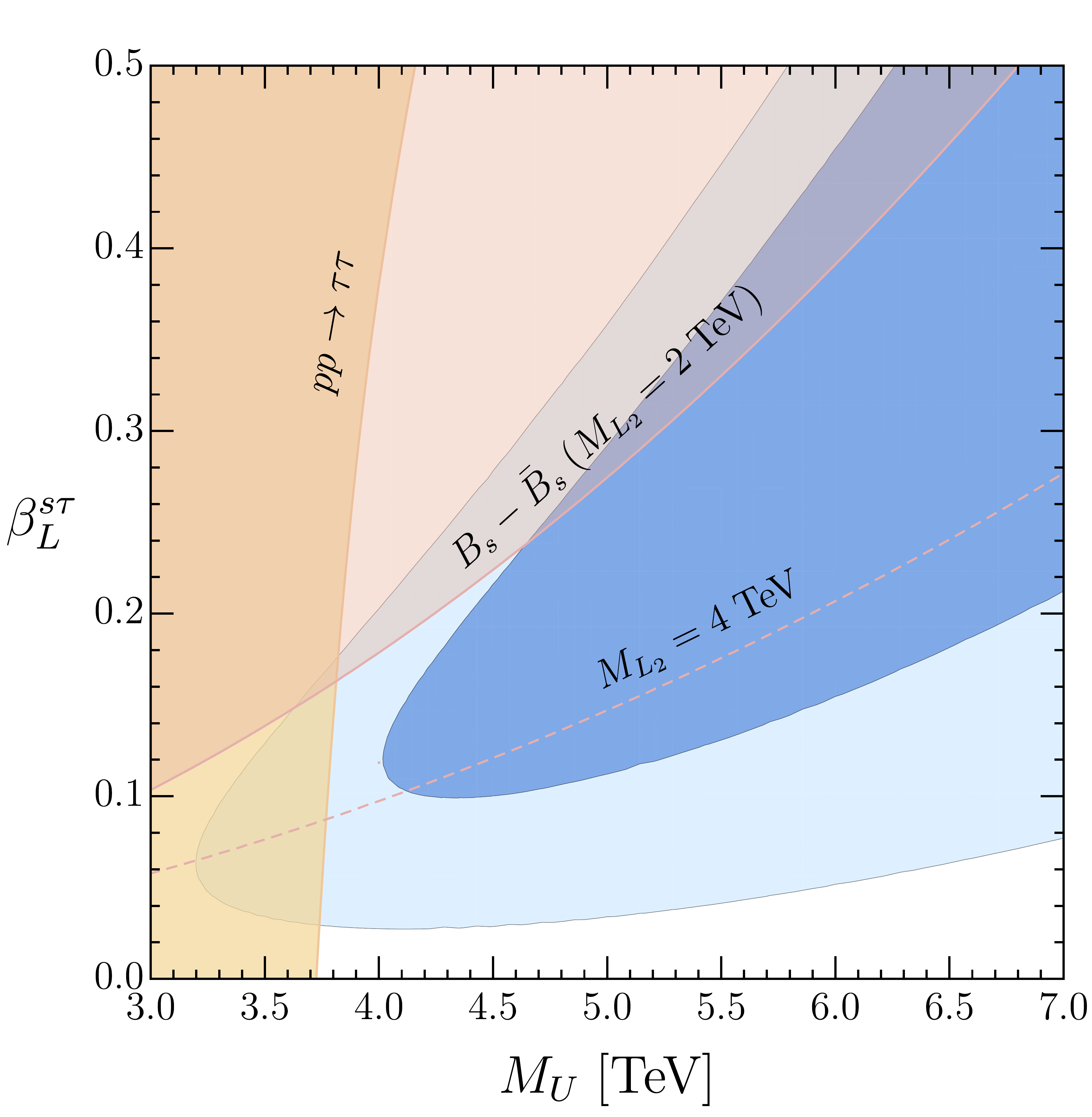}
~~~~~~~~~~~ 
\includegraphics[width=0.45\textwidth]{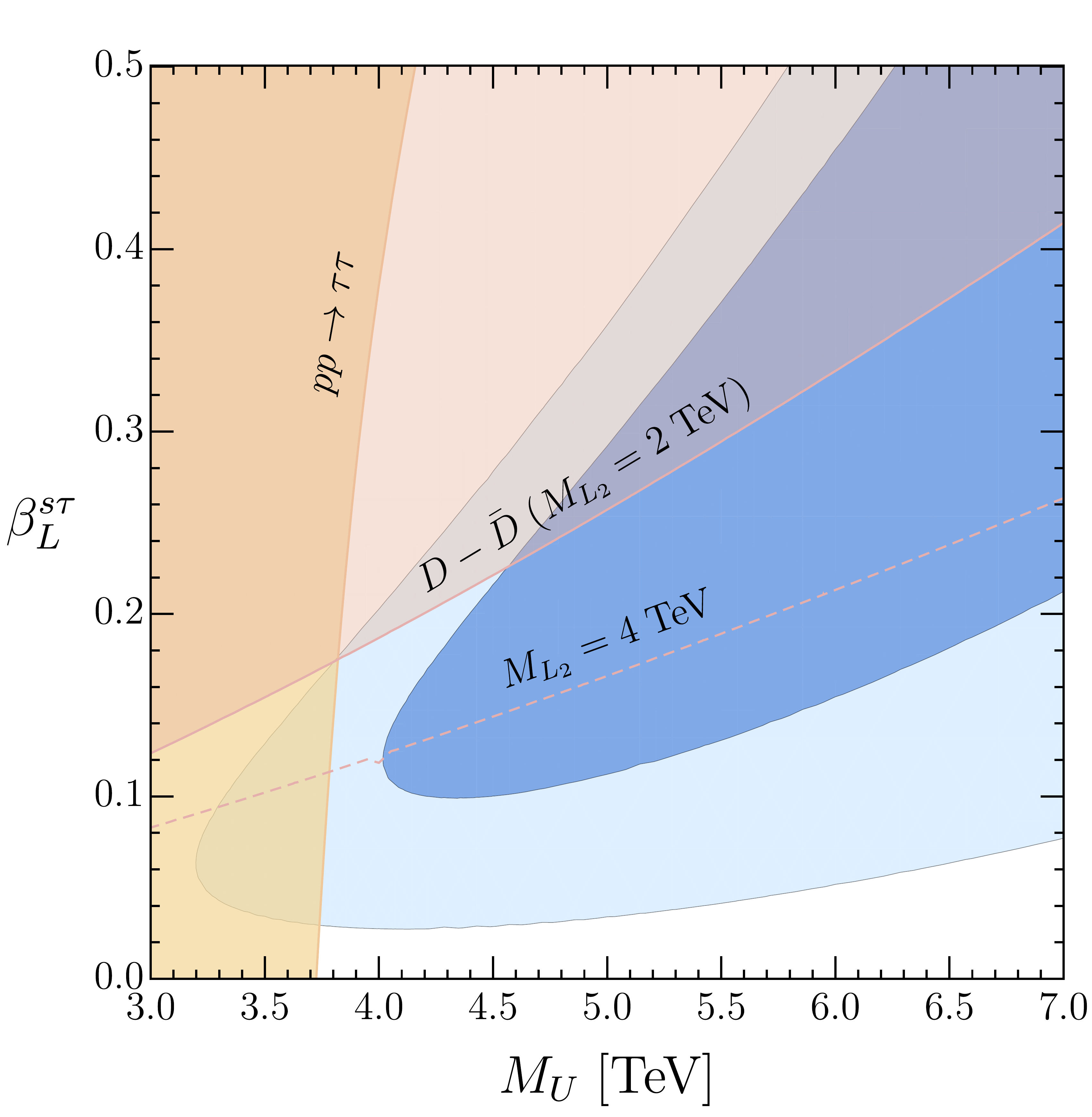} 
\caption{$95\%$~CL constraints from $B_s-\bar B_s$ and $D-\bar D$ for different benchmarks values of the vector-like lepton mass $M_{L_2}$ and with $g_U=3.0$. The $1\sigma$ and $2\sigma$ regions preferred by the low-energy fit are shown in blue and light blue, respectively. For reference, we also show the high-$p_T$ bound from $pp\to\tau\tau$ taken from~\cite{Baker:2019sli} and discussed in Section~\ref{ssec:highpT}. In the right plot we fix $s_\chi=0.55$ and $s_{\ell_2}=0.15$.}
\label{fig:MesonMix}
\end{figure}

We parametrize the $U_1$ contributions to $\Delta F=2$ observables by the following effective Lagrangians
\begin{align}\label{DF2Lag}
\mathcal{L}_{\Delta B=2} &= C_{B_i}\,\left(\bar b_L\gamma_\mu d^i_L\right)^2\,,&&&
\mathcal{L}_{\Delta S=2}&=C_K\,\left(\bar s_L\gamma_\mu d_L\right)^2\,, &&&
\mathcal{L}_{\Delta C=2}& =C_D\,\left(\bar c_L\gamma_\mu u_L\right)^2\,.
\end{align}
The SM contribution to the $\Delta B=2$ operator reads
\begin{align}
C_{B_i}^{\rm SM}(m_b)\approx\frac{G_F^2 M_W^2}{4\pi^2}\,(V_{tb}^*V_{ti})^2\,S_0(x_t)\,\eta_B\,,
\end{align}
with $\eta_B$ being a running factor, $S_0(x)$ the Inami-Lim function~\cite{Inami:1980fz} and $x_t=m_t^2/M_W^2$. NP contributions to this operator mediated by the $U_1$ at one loop have been computed in~\cite{DiLuzio:2018zxy}. The same result applies also to our model, given that the $U_1$ right-handed couplings in~\eqref{eq:U1coup} do not contribute to this observable. We have
\begin{align}\label{eq:DB=2}
C_{B_i}^U(m_b)=-\frac{C_U^2 M_U^2\, G_F^2}{4\pi^2}\,\eta_U\sum_{\ell,\ell^\prime}\lambda_{B_i}^\ell\lambda_{B_i}^{\ell^\prime}\,F(x_\ell,x_{\ell^\prime})\,,
\end{align}
where $\eta_U$ accounts for the running from $M_U$ to $m_b$, $\lambda_{B_i}^\ell=\beta_L^{b\ell}\,(\beta_L^{i\ell})^*$, $x_\ell=M_\ell^2/M_U^2$ (with $M_\ell$ the mass of the lepton running in the loop) and the loop function $F(x_\ell,x_{\ell^\prime})$ can be found in~\cite{DiLuzio:2018zxy}. In the evaluation of~\eqref{eq:DB=2} we have removed $x_\ell$-independent terms, which cancel after using the unitarity of the fermion-mixing matrices. The final result is finite only after all the fermions entering in the loop, including the vector-like leptons, are included. The dominant NP contribution is given by the most massive particle in the loop, in this case one of the vector-like leptons. Neglecting the SM lepton masses, we find
\begin{align}\label{eq:Cbs1}
C_{B_i}(m_b)\equiv1+\frac{C_{B_i}^U(m_b)}{C_{B_i}^{\rm SM}(m_b)}&\approx 1+\frac{C_U^2 M_U^2}{M_W^2}\left(\frac{\beta_L^{bL_2}\beta_L^{iL_2\,*}}{V_{tb}^*V_{ti}}\right)^2\,\frac{S_0(x_{L_2})}{S_0(x_t)}\,\frac{\eta_U}{\eta_B}\,.
\end{align}
Note that, due to the flavor structure in~\eqref{eq:U1coup}, we have $C_{B_d}^U\approx0$, while the $U_1$ contribution to $B_s$-mixing can be sizable. Taking $c_{\ell_2}\approx1$, we can use the relation $\beta_L^{bL_2}\beta_L^{sL_2}\approx -\beta_L^{b\tau}\beta_L^{s\tau}/c_d$ to write the $B_s$-mixing contribution in terms of the parameters used in Section~\ref{sec:LQDyn}. Taking the bounds on $C_{B_s}(m_b)$ from $\Delta m_s$ provided by UTfit~\cite{Bona:2017cxr}, we extract an upper limit on $M_{L_2}$ of a few TeV, see Figure~\ref{fig:MesonMix}. 

It should be stressed that the growth of $\Delta F=2$ amplitudes with the vector-like mass is an artifact due to our choice 
of expressing the result in terms of the $\beta_L^{ij}$ couplings. Indeed  working in the limit of small mixing,
in analogy to Eq.~(\ref{eq:bst_approx}), $C_{B_s}$ can be expressed as follows
\begin{equation}\label{eq:Cbs2}
C_{B_s}(m_b)   \approx  1 +  \frac{1}{24}  \frac{\eta_U}{\eta_B} \frac{v^2}{M_Q^2}  \left[  \frac{g_U^4 \omega_3^2 \omega_{15}^2}{M_U^4} \right]
\frac{  \lambda^2_{15} \lambda^2_{q} }{ y_t^2\, V_{ts}^2}\, (1 +\rho)\,,
\end{equation}
where $y_t$ is the top-quark Yukawa coupling and $\rho$ is an $\mathcal{O}(1)$ term depending on the details of the spectrum.
This results exhibits the expected decoupling behavior with the NP masses and the power growth 
with the symmetry breaking parameters  $\lambda_{15}$ and $\lambda_{q}$. 
From Eq.~(\ref{eq:Cbs2}) it is evident that the $B_s$-mixing constrains the maximal size of  $\lambda_{15}$ and $\lambda_{q}$. 
On the other hand, if we wish to keep the couplings fixed (in particular $\beta_L^{s\tau}$) given the 
information derived from the low-energy fit, then the $B_s$-mixing bound   can be translated into an upper bound on the  vector-like masses
(as shown in Figure~\ref{fig:MesonMix}).  From this point of view the situation resembles the SM case, where the charm quark 
was predicted in order to render the SM loop contribution finite~\cite{Glashow:1970gm}, and a rough estimation of the charm mass was obtained from $K-\bar K$ mixing~\cite{Gaillard:1974hs}.

Proceeding in a similar way with $D-\bar D$ mixing 
we find\footnote{Contrary to the $\Delta B=2$ case, we have tree-level contributions to $\Delta C=2$ transitions, meditated by the $Z^\prime$ and $G^\prime$. In the $U(2)_q$-preserving limit, these contributions are proportional to $(V_{ub}^*V_{cb})^2$ and therefore negligibly small~\cite{DiLuzio:2018zxy}. In our loop calculation, we consistently remove terms of $\mathcal{O}[(V_{ub}^*V_{cb})^2]$, which are negligible compared to their tree-level counterparts.}
\begin{align}
C_{D}^U & =-\frac{C_U^2 M_U^2\, G_F^2}{4\pi^2}\,\sum_{\ell,\ell^\prime}\lambda_{D}^\ell \lambda_{D}^{\ell^\prime}\,F(x_\ell,x_{\ell^\prime})\,,
\end{align}
where $\lambda_D^\ell=V_{ui}^*\,V_{cj}\,(\beta_L^{i\ell})^*\,\beta_L^{j\ell}$. We get important constraints from NP contributions to the imaginary part of $C_D^U$. We can interpret these constraints as a bound on $\beta_L^{s\tau}$ as a function of $M_U$, once we fix the vector-like lepton masses, $s_\chi$ and $s_{\ell_2}$.  This is shown in~Figure~\ref{fig:MesonMix} where we use the latest UTfit analysis for the $D-\bar D$ constraint~\cite{Silvestrini:LaThuile,Carrasco:2014uya}. On the other hand, as it happens with $B_d$-mixing, the contributions to $K-\bar K$ mixing are suppressed by the SM lepton masses and are thus negligible.

\subsubsection{Dipole contributions} 
It was noted in~\cite{Crivellin:2018yvo} that a large $\beta_L^{s \tau}$ could yield sizable $b\to s$ dipole contributions mediated by the $U_1$ at one loop.  Recasting the results in~\cite{He:2009rz,Konig:2014iqa} and neglecting terms proportional to $m_s$, we find ($\ell = \mu, \tau, L_2$)
\begin{align}\label{eq:C7}
\begin{aligned}
\Delta \mathcal{C}_{7 (8)} &=  \frac{C_{U}}{V_{tb}V_{ts}^{*}}   \sum_{\ell}   \beta_L^{s \ell}  (\beta_L^{b \ell})^*  \left( \frac{x_\ell \left(4 + 25 x_\ell + x_\ell^2\right)}{24 (1-x_\ell)^3} +\frac{x_\ell^2 (3 + 2 x_\ell) \log x_\ell}{4(1-x_\ell)^4}\right) \\
& \quad + \beta_L^{s \tau} (\beta_R^{b \tau})^*\, \frac{m_{\tau}}{m_{b}} \left( \frac{4 + 25 x_{\tau} +  x_{\tau}^{2} }{12( 1- x_{\tau})^{2}} +\frac{x_\tau (3 + 2 x_\tau) \log x_\tau}{2(1-x_\tau)^3} \right) \,,
\end{aligned}
\end{align}
and $\Delta \mathcal{C}_{7^\prime,8^\prime}\approx0$ (see Appendix~\ref{app:EffHam} for the Wilson coefficient definitions). For $\beta_R^{b\tau}=-1$ we find that the $LR$ term gives the dominant contribution. Using the low-energy fit values from Section~\ref{ssec:fit}, and taking into account the running from the TeV scale to $m_{b}$~\cite{Konig:2014iqa}, we get $\Delta \mathcal{C}_{7,8}(m_b) \sim \mathcal{O}(10^{-3})$, well below the current bounds~\cite{Capdevila:2017bsm}. We find that the $LL$ contribution is smaller by two orders of magnitude compared to the one found in~\cite{Crivellin:2018yvo}. This difference is due to the cancellation of the $x_\ell$-independent terms proportional to  $\beta_L^{s \tau}  (\beta_L^{b \tau})^*$, once we also include the vector-like lepton in the loop. Once more, we note the importance of computing these loops in a UV-complete model.

The dominant (chirally enhanced) contribution to $\tau\to\mu\gamma$ was already computed in the dynamical model in Section~\ref{sec:LQDyn}. The full $U_1$ contribution, now calculable in the UV-complete model, is found to be~\cite{He:2009rz} ($q=s,b,Q_1,Q_2$)
\begin{align}
\begin{aligned}
\mathcal{B}(\tau \to \mu \gamma) & =   \frac{1}{\Gamma_{\tau}} \frac{\alpha}{4096 \pi^{4}}       \frac{m_{\tau}^3 m_{b}^{2}  }{ v^{4}} \, C_{U}^{2}   \left| 2\,  \beta_L^{b \mu} (\beta_R^{b \tau})^* \left(\frac{4 - 23 x_b + x_b^2}{(1 - x_b)^2} - \frac{6 x_b (1 + 2 x_b) \log x_b}{(1 - x_b)^3} \right)  \right. \\
&\quad  \left. - \frac{m_{\tau}}{  m_{b} }\, \sum_q\,\beta_L^{q \mu} (\beta_L^{q \tau})^*  \left(  \frac{3 x_q^2 (5+x_q)}{(1 - x_q)^3}  +  \frac{6 x_q^2 (1 + 2 x_q) \log x_q}{(1 - x_q)^4} \right)  \right|^{2} \,,
\end{aligned}
\end{align}
where we have ignored terms proportional to $m_\mu$ and, as in the previous computations, we have used the unitarity of the fermion-mixing matrices to remove the $x_q$-independent terms in the $LL$ contribution. We have explicitly checked that when $\beta_R^{b \tau}=-1$ the $LL$ contributions are much smaller than the ones included in~\eqref{eq:tau2mugamma}, which justifies having neglected them in the low-energy fit.

%%%%%%%%%%%%%%%%%%%%%%%%%%%%%%%%
\subsection{Constraints on the new fields}\label{sec:constraints}
%%%%%%%%%%%%%%%%%%%%%%%%%%%%%%%%
The UV-complete model introduced in Section~\ref{sec:model} contains additional fields beyond the  $U_1$ that could potentially alter some of the results obtained in Section~\ref{sec:LQDyn}. In what follows, we discuss the main constraints on these particles
\begin{itemize} 
\item \textbf{Additional vectors.} 
If we assume perfect alignment to down-type quarks, as we did in~\eqref{eq:U1coup}, 
the $Z^\prime$ and $G^\prime$ couplings to fermions are given in Appendix~\ref{app:VecCoup}. In this limit, the only $\Delta F=2$ amplitude 
receiving relevant tree-level contributions from  $Z^\prime$ and $G^\prime$ exchange is $D-\bar D$ mixing.\footnote{A small tree-level effect is also generated in the $K-\bar K$ mixing amplitude. Given its smallness and the fact that it mostly contributes to the real part of the mixing amplitude, this effect is unobservable.} 
Using the same notation as in \eqref{DF2Lag} for the Wilson coefficient and taking the mixing angles in \eqref{qlmix}, we can write the contribution to $\Delta C=2$ transitions as (see also~\cite{Bordone:2018nbg,DiLuzio:2018zxy})
\begin{align}\label{eq:C1D_tree}
\begin{aligned}
\left.C_1^D\right|_{\rm tree}&\approx\frac{4 G_F}{ \sqrt{2} }\left(C_{Z^\prime}+\frac{C_{G^\prime}}{3}\right)\,(V_{ub}^*\,V_{cb})^2\,\left(1-s_{q_1}^2-c_d^2\,s_{q_1}^2 s_{\chi_q}^2 \, \left|\frac{V_{tb}}{V_{ts}}\right|^2\right)^2\,,
\end{aligned}
\end{align}
where $C_{Z^\prime}$ and $C_{G^\prime}$ are given by   
\begin{align}\label{eq:Cs}
C_{Z^\prime}&=\frac{g_Y^2}{24\,g_1^2}\,\frac{g_4^2v^2}{4\,M_{Z^\prime}^2}\,, & C_{G^\prime}&=\frac{g_c^2}{g_3^2}\,\frac{g_4^2v^2}{4\,M_{G^\prime}^2}\,.
\end{align}
In the $U(2)_q$-preserving limit, i.e. when $s_{q_1}s_{\chi_q}=0$, these contributions are strongly CKM-suppressed and the net effect on $C_1^D$ is of $\mathcal{O}(10^{-9})~\mathrm{TeV}^{-2}$ for both real and imaginary parts, well compatible with the current limits from UTFit~\cite{Silvestrini:LaThuile,Carrasco:2014uya}. On the other hand $U(2)_q$-breaking effects, parametrized by $s_{q_1}s_{\chi_q}$, are CKM-enhanced compared with the latter contribution and could be potentially dangerous. Using typical values for the model parameters, we estimate that the  $U(2)_q$-breaking term can be as large as $|s_{q_1}s_{\chi_q}|\approx0.07$, while remaining consistent with the $D-\bar D$ constraint. Using the relations in~\eqref{eq:chiMix}, we find that it is possible to obtain sizable values for $\beta_L^{s\tau}$, as required by the low-energy fit, while keeping the NP contributions to $D-\bar D$ well below the current bounds.

The additional vectors are in the interesting range for high-$p_T$ searches at LHC. The related collider signatures have been extensively analyzed in general terms in~\cite{Baker:2019sli}. Here we comment on the main implications for the benchmark $g_4=3.0$ (implying $g_4\gg g_3\gg g_1$), for which the $Z^\prime$ and $G^\prime$ interactions to light-generation quarks and leptons are suppressed. The most important constraint on the $G^\prime$ is obtained from $pp\to tt$, which sets a lower limit on its mass of $M_{G^\prime}\gtrsim 3.5$~TeV. Given the mass relation between the $U_1$ and $G^\prime$ (see~\eqref{eq:VectorMasses} taking $g_4\gg g_3$),
\begin{align}\label{eq:GpUrel}
M_{G^\prime}&\approx M_U\,\sqrt{\frac{2\,\omega_3^2}{\omega_1^2+\omega_3^2+\frac{4}{3}\omega_{15}^2}}\,, 
\end{align}
we find that current high-$p_T$ bounds on the $G^\prime$ are typically less constraining (although comparable) than the ones on the $U_1$ for most of the parameter space. This is in contrast to other UV completions where the vector leptoquark only couples to left-handed fermions, as e.g. in~\cite{DiLuzio:2018zxy}. The most relevant channel for direct searches on the $Z^\prime$ is $pp\to\tau\tau$, from which we obtain a mass limit of $M_{Z^\prime}\gtrsim2.5$. The $Z^\prime$ contributions to this channel could potentially affect the discussion in Section~\ref{ssec:highpT}. However, these contributions drop fast with increasing $Z^\prime$ mass and become negligible once $M_{Z^\prime}\gtrsim 3.0$~TeV.

\item \textbf{Vector-like fermions.} Vector-like fermions are predicted to be among the lightest new states in the model. High-$p_T$ searches involving these particles therefore constitute a very interesting probe for the proposed scenario. Most of the results obtained in~\cite{DiLuzio:2018zxy} apply also to our model. However, in our case the vector resonances and vector-like fermions are heavier, resulting in typically smaller production cross sections. As shown in Section~\ref{sec:DF2}, vector-like lepton masses are expected to lie around 2--4~TeV. A mass splitting between vector-like quarks and leptons is generated after $\Omega_{15}$ takes a vev (see~\eqref{eq:Mchisplit}), resulting in vector-like quarks masses that are around one TeV larger than the ones of the vector-like leptons. 

As in~\cite{DiLuzio:2018zxy}, the dominant production mechanism for the vector-like quarks is not via QCD interactions but via the $G^\prime$ through the processes $q\bar q\to G^\prime\to Q\bar Q, Q\bar q, q\bar Q$. The $G^\prime$ dominantly decays to vector-like pairs while the SM-vector-like combination is suppressed by one power of $s_{q_{1,2}}$. Vector-like leptons are produced via electroweak interactions. Neutral current processes receive additional contributions from $Z^\prime$-assisted production which is stronger than the eletroweak production by more than one order of magnitude. Analogously to the vector-like quark case, mixed $Z^\prime$ decays involving a SM and a vector-like lepton are suppressed by one power of $s_{\ell_2}$. We implement the model in {\tt FeynRules}~\cite{Alloul:2013bka} and use {\tt Madgraph5\_aMC@NLO}~\cite{Alwall:2014hca} to compute the production cross-sections. We find that the production cross-sections for both vector-like quarks and leptons are well below $1~\mathrm{fb}$ in the relevant range of model parameters, and therefore out of the LHC reach. 

Second-family vector-like fermions can have sizable couplings to the Higgs, and they are expected to decay dominantly to a third-generation SM fermion of the same type and a $h$, $W$ or $Z$. Current limits on pair-produced vector-like quarks and leptons with these decay channels are of $\mathcal{O}(10)~\mathrm{fb}$~\cite{Aaboud:2018pii,CMS:2018cgi}. The situation is different for the first-family vector-like fermions. As in~\cite{DiLuzio:2018zxy}, their coupling to the Higgs are extremely suppressed by the first-generation fermion masses, so they are expected to decay predominantly to three third-generation SM fermions via an off-shell heavy vector.\footnote{This decay channel can also dominate over the two-body decay for the second-generation vector-like fermions whose couplings to the Higgs are accidentally suppressed. This is for instance the case for the down-type vector-like quark if we assume perfect alignment to SM down-type quarks, as we did in~\eqref{eq:U1coup}.} In this case, the vector-like signatures in the detector contain multiple jets and leptons and are rich with $b$-tags and $\tau$-tags. While a dedicated analysis on these signatures is needed, one can extract a rough estimate on the production cross-section by comparing with existing supersymmetry searches~\cite{Aaboud:2017dmy}. The limit found in~\cite{DiLuzio:2018zxy} is around $5-15~\mathrm{fb}$, depending on the decay topologies.

\item \textbf{Additional scalars.} A dedicated analysis of the scalar sector of the model is beyond the scope of this paper (a detailed analysis for a very similar setup can be found in~\cite{DiLuzio:2018zxy}). The masses of the additional scalars depend on the scalar potential parameters, which are mostly unconstrained, but they are expected to be around a few TeV. The Yukawa couplings of the radial excitations in $\Omega_{1,3,15}$ necessarily involve a SM and a vector-like fermion, see~\eqref{eq:O13Chi} and \eqref{eq:O15Chi}. Therefore, they can only affect low-energy observables at the one-loop level. Moreover, flavor-changing transitions are protected by the same flavor structure that controls the vector boson interactions. As a result, we conclude that these scalars do not yield relevant effects at low energies. Apart from the additional Higgs doublet, the $H_{15}$  also features a $R_2$ and a $\tilde R_2$ leptoquarks and a color octet charged under $SU(2)_L$. These scalars have Yukawa interactions with two SM fermions and could potentially mediate relevant low-energy effects. Also in this case, the Yukawa interactions present the same flavor structure discussed in~\ref{sec:mixing}: dominant couplings to third-generation fermions, with small couplings to left-handed light-generation fermions and negligible couplings to right-handed light-generation fermions. The $R_2$ leptoquark was recently proposed as a solution to the $R_{D^{(*)}}$ anomalies~\cite{Becirevic:2018afm}. However in our model the $R_2$ contributions to these observables are negligible due to the smallness of the light-generation right-handed couplings. On the other hand, the $R_2$ leptoquark could yield potentially large contributions to $\mathcal{B}(\tau\to\mu\gamma)$ at the one-loop level which are chiral enhanced by a factor $m_t/m_\tau$, see e.g.~\cite{Becirevic:2017jtw}. We find that the $R_2$ contributions to this observable are below the current bounds provided its Yukawa interactions are of $\mathcal{O}(10^{-1})$, for a mass of $2$~TeV. The $\tilde R_2$ leptoquark has also been proposed as a possible explanation of the $R_{D^{(*)}}$ anomalies if one introduces a light $\nu_R$ that fakes the SM ones, as e.g. in~\cite{Becirevic:2016yqi,Robinson:2018gza}. We will not consider this possibility here. 

Concerning direct searches, the most interesting states to look for at high-$p_T$ are the colored ones, since they can be produced via QCD interactions. Following the discussion in~\cite{DiLuzio:2018zxy} (see also~\cite{Bai:2018jsr}), we conclude that the production cross-sections of the radial modes in $\Omega_{1,3,15}$ are small enough to avoid detection at the LHC provided their masses are around a few TeV. Similar conclusions also hold for the charged color-octect and the $R_2$ and $\tilde R_2$ leptoquarks~\cite{Angelescu:2018tyl,Marzocca:2018wcf,Faber:2018afz}.

\end{itemize}

\noindent 
We therefore conclude that the presence of the additional particles does not affect the phenomenological implications of the $U_1$ 
derived in terms of the simplified model. 
However, the UV-complete model presents many interesting signatures that go beyond the simplified setup and 
whose exploration could be an essential ingredient to test the $U_1$ solution of the B-anomalies and possibly reconstruct the underlying NP model.

\bigskip
For the sake of completeness, we report here a 
benchmark point that provides a good low-energy fit and satisfies the high-$p_T$ constraints discussed 
in Section~\ref{sec:LQDyn}, as well as the additional low-energy constrains discussed above:
\begin{align}
\begin{aligned}
g_4&= 3.0\, , 
&&&  \omega_1&= 1.0~\mathrm{TeV}\,, &&& \omega_3&= 2.2~\mathrm{TeV}\,, &&& \omega_{15}&=1.5~\mathrm{TeV}\,, &&& M_\chi&= 3.0~\mathrm{TeV}\,,  \\
\lambda_\ell &= 0.25\,,   &&& \lambda_q&= 0.25\,, &&& \lambda_{15}&=-1.2\,, &&& \lambda_{15}^\prime&= 1.0\,,    &&& s_\tau&=0.05\,.
\label{eq:Yukawas_BM}
\end{aligned}
\end{align}
From these values we obtain the following spectrum
\begin{align}
\begin{aligned}
M_U&= 4.5~\mathrm{TeV}\,, &&& M_{Z^\prime}&= 3.5~\mathrm{TeV}\,, &&& M_{G^\prime}&=5.0~\mathrm{TeV}\,,\\
M_{Q_1} \approx M_{Q_2} &= 3.3~\mathrm{TeV}\,, &&& M_{L_1}&=2.1~\mathrm{TeV}\,, &&& M_{L_2}&= 2.3~\mathrm{TeV}\,, \\
\end{aligned}
\end{align}
and mixing angles  $\{ s_{\ell_2}, s_{q_1} , s_\chi, s_{\chi_q},s_{\chi_\ell}\}=\{ 0.12, 0.21, 0.55, -0.11, 0.46\}$,
resulting in  the following effective leptoquark couplings: $\{\beta_R^{b\tau},\beta_L^{b\tau},\beta_L^{s\tau},\beta_L^{b\mu},\beta_L^{s\mu},\beta_L^{d\tau}\}=\{-1.0, 0.84, 0.11,-0.11, 0.02, -0.02\}$. 
This benchmark point should not be considered as a particularly favored configuration. It is only 
an illustration that is possible to reach the allowed region 
of the spectrum consistent with data (Figures~\ref{fig:highpT_pheno} and \ref{fig:MesonMix}),
as well as the $U_1$ couplings identified by the low-energy fit (Figure~\ref{fig:2Dfit}), 
with very reasonable Lagrangian parameters. We stress in particular the 
smallness of the Yukawa couplings in~\eqref{eq:Yukawas_BM}, which do not raise 
perturbativity issues up to very high energy scales. The only tuning of the model is the 
ansatz in~\eqref{eq:lambda_hat} for the $U(2)_q\times U(2)_\ell$ flavor symmetry breaking terms,
and their alignment to the down-type quark and charged-lepton mass-eigenstate basis 
(signaled by the smallness of $s_b$ and $s_\tau$). However, these are radiatively stable 
conditions that can be enforced via suitable dynamical constraints
on the symmetry-breaking fields.

%%%%%%%%%%%%%%%%%%%%%%%%%%%%%%%%%%%%%%%%%%%%%%%%%%%%%%%%%%%%%%%%%%
\section{Conclusions}\label{sec:conclusions}
%%%%%%%%%%%%%%%%%%%%%%%%%%%%%%%%%%%%%%%%%%%%%%%%%%%%%%%%%%%%%%%%%%

Among the different options proposed to explain the hints of LFU violation observed in $B$-meson decays,
the hypothesis of a $SU(2)_L$-singlet vector leptoquark ($U_1$)
stands for its simplicity and effectiveness.
In this paper we have presented a thorough investigation of this hypothesis 
from a twofold perspective: first using a simplified-model/EFT approach, taking into account 
recent results from $B$-physics observables 
and high-$p_T$ searches,  and then presenting 
a more complete model with a consistent UV behavior. 

Employing the simplified model we have shown that a right-handed coupling 
for the $U_1$, mainly aligned to the third-generation, can be a virtue rather than a problem.
This coupling, neglected in most previous studies,
can yield a very good fit of the $b\to c$ anomalies without significant drawbacks. 
The outcome of the low-energy fit with the inclusion of right-handed couplings has
been presented in Section~\ref{ssec:fit}. A key difference with respect to
previous studies is the strong enhancement (compared to the SM predictions)
of the rates for helicity-suppressed modes with tau leptons,  
in particular $B_s \to \tau^+\tau^-$ and $B_s \to \tau \mu$.
The experimental search of these decays modes, whose expectation is not far
from present bounds, could provide a smoking-gun signature for 
this framework (or could lead us to rule it out). An additional important implication of the right-handed coupling for the $U_1$ 
is the larger impact on $b\to c$ anomalies at fixed $U_1$ mass. This fact, together with
the reduced deviation from the SM indicated by the recent Belle analysis~\cite{Belle:Moriond}, leads to an excellent 
compatibility between low- and high-energy data in this framework, at least at present. 
Interestingly enough, the preferred mass--coupling range for the $U_1$ inferred by the anomalies is
well within the reach of direct searches at the HL-LHC (see Figure~\ref{fig:highpT_pheno}).

In the second part of the paper we have shown how a simple extension of the matter content of the model 
proposed in~\cite{Bordone:2017bld},  based on a flavor deconstruction of the original Pati-Salam gauge group, 
provides a good UV completion for the $U_1$ with the precise couplings to SM fermions required to describe 
current data. The field content of the model
is summarized in Table~\ref{tab:fieldcontent}.
The most important consequence following from the requirement of a consistent UV completion
 is the necessity of extra TeV scale fields, with interesting high-$p_T$ signatures 
that cannot be deduced within the simplified model. 
These new states include both a color-octet ($G^\prime$) and a color singlet ($Z^\prime$) vector field, 
as extensively discussed in \cite{DiLuzio:2017vat,Bordone:2017bld,Bordone:2018nbg,Greljo:2018tuh}, 
and a pair of vector-like quarks and leptons. As we have shown, and as already pointed out in \cite{DiLuzio:2018zxy}, 
the $\Delta F=2$ constraints imply that the vector-like leptons are likely to be the lightest exotic states. 

In conclusion, our analysis reinforces the phenomenological success of the vector leptoquark hypothesis 
in explaining the hints of LFU violation observed in $B$-meson decays, taking into account all 
available low- and high-energy data. We also confirm the compatibility of this hypothesis with
motivated extensions of the SM based on the idea of flavor non-universal gauge interactions~\cite{Bordone:2017bld}, 
which could provide an explanation for the long-standing puzzle of quark and lepton masses.

%%%%%%%%%%%%%%%%%%%%%%%%%%%%%%%%%%%%%%%%%%%%%%%%%%%%%%%%%%%%%%%%%%
\section*{Acknowledgements}
%%%%%%%%%%%%%%%%%%%%%%%%%%%%%%%%%%%%%%%%%%%%%%%%%%%%%%%%%%%%%%%%%%

We thank Riccardo Barbieri for interesting discussions and 
we are grateful to Joaquim Matias and Javier Virto for providing us their results for the fit to $b\to s\ell\ell$ data prior to publication.
This research was supported in part by the Swiss National Science Foundation (SNF) under contract 200021-159720.

%%%%%%%%%%%%%%%%%%%%%%%%%%%%%%%%%%%%%%%%%%%%%%%%%%%%%%%%%%%%%%%%%%
\appendix
%%%%%%%%%%%%%%%%%%%%%%%%%%%%%%%%%%%%%%%%%%%%%%%%%%%%%%%%%%%%%%%%%%

%%%%%%%%%%%%%%%%%%%%%%%%%%%%%%%%
\section{The Weak Effective Hamiltonian}\label{app:EffHam}
%%%%%%%%%%%%%%%%%%%%%%%%%%%%%%%%
Semileptonic and dipole $b\to s$ transitions are commonly parameterized in terms of the so-called Weak Effective Theory (WET)~\cite{Grinstein:1987vj,Buchalla:1995vs,Buras:1998raa} 
\begin{equation}
\mathcal{H}_{\rm WET} \supset - \frac{4G_F}{\sqrt2}\,\frac{e^2}{16\pi^2}\,V_{tb}V_{ts}^* \sum_i \Big[ \mathcal{C}_i\,\mathcal{O}_i + h.c. \Big]\,,
\label{HWET}
\end{equation}
where the operators
\begin{align}\label{eq:WET_op}
\begin{aligned}
\mathcal{O}^{\ell}_9&=\left(\overline{s}\gamma_\mu P_Lb\right)\left(\overline{\ell}\gamma^\mu \ell\right)\,,    &\quad&& \mathcal{O}_{9^\prime}^\ell &=  \left(\overline{s}\gamma_\mu P_Rb\right)\left(\overline{\ell}\gamma^\mu \ell\right)\,,\\[5pt] 
 \mathcal{O}^{\ell}_{10}&= \left(\overline{s}\gamma_\mu P_Lb\right)\left(\overline{\ell}\gamma^\mu\gamma_5 \ell\right)\,,   &&& \mathcal{O}_{10^\prime}^\ell&= \left(\overline{s}\gamma_\mu P_Rb\right)\left(\overline{\ell}\gamma^\mu\gamma_5 \ell\right)\,,    \\[5pt]
\mathcal{O}^{\ell}_{S}&= m_b (  \bar s P_R b )( \bar \ell \ell) \,,  &&&\mathcal{O}^\ell_{S^\prime}&= m_b (  \bar s P_L b )( \bar \ell \ell) \,,  \\[5pt] 
\mathcal{O}^{\ell}_{P}&= m_b (  \bar s P_R b )( \bar \ell  \gamma_5 \ell) \,,  &&&\mathcal{O}^\ell_{P^\prime}&= m_b (  \bar s P_L b )( \bar \ell  \gamma_5 \ell) \,,\\[5pt]
 \mathcal{O}_{7} &= \frac{m_b}{e}\,  ( \bar{s}  \, \sigma_{\mu\nu}\, P_R  \,  b ) \; F^{\mu\nu},&&& \mathcal{O}_{7^\prime}&= \frac{m_b}{e}\,  ( \bar{s}  \, \sigma_{\mu\nu}\, P_L \,  b ) \; F^{\mu\nu},\\[5pt]
  \mathcal{O}_{8} &= \frac{g_c\,m_b}{e^2}\,  ( \bar{s}  \, \sigma_{\mu\nu}\, P_R\,T^a  \,  b ) \; G^{\mu\nu\,a},&&& \mathcal{O}_{8^\prime}&= \frac{g_c\,m_b}{e^2}\,  ( \bar{s}  \, \sigma_{\mu\nu}\,P_L\, T^a \,  b ) \; G^{\mu\nu\,a}\,,
\end{aligned}
\end{align}
with $\ell=e,\mu,\tau$ and $P_{L,R}=1/2(1\mp\gamma_5)$. The corresponding Wilson coefficients are parametrized as $\mathcal{C}_i^\ell=\mathcal{C}_i^{\rm SM}+\Delta\mathcal{C}_i^\ell$, where $\mathcal{C}_i^{\rm SM}$ denotes the SM contribution and $\Delta\mathcal{C}_i^\ell$ encodes possible NP effects.

%%%%%%%%%%%%%%%%%%%%%%%%%%%%%%%%
\section{$Z^\prime$ and $G^{\prime}$ couplings to fermions}\label{app:VecCoup}
%%%%%%%%%%%%%%%%%%%%%%%%%%%%%%%%
For completeness, in this section we provide the $Z^\prime$ and $G^\prime$ couplings to fermions in their mass eigenbasis. Collecting the left-handed fermions in 5-dimensional multiplets, as in~\eqref{eq:ULag}, we obtain 
\begin{align}
\begin{aligned}
\mathcal{L}_{G^{\prime}}&\supset g_c\,\frac{g_4}{g_3}\,G^{\prime \, a}_\mu\left[\kappa_q\,(\bar \Psi_q\gamma^\mu\,T^a\, \Psi_q)+\kappa_u\,(\bar u_R \gamma_{\mu}\,T^a\, u_R) + \kappa_d\,(\bar d_R \gamma_{\mu}\,T^a\, d_R) + \kappa_Q\,(\bar Q_R \gamma_{\mu}\,T^a\, Q_R) \right]\,,\\
\mathcal{L}_{Z^\prime}&\supset \frac{g_Y}{2\sqrt{6}}\,\frac{g_4}{g_1}\,Z_\mu^\prime\left[\xi_q\,(\bar \Psi_q\gamma^\mu \Psi_q)+\xi_u\,(\bar u_R\gamma^\mu u_R)+\xi_d\,(\bar d_R\gamma^\mu d_R) + \xi_Q\, (\bar Q_R \gamma_{\mu} Q_R) -3\,\xi_\ell\,(\bar \Psi_\ell\gamma^\mu \Psi_\ell) \right. \\
&\quad \left. -3\,\xi_e\,(\bar e_R\gamma^\mu e_R) -3\,\xi_L\, (\bar L_R \gamma_{\mu} L_R) \right]\,.
\end{aligned}
\end{align}
Using the same flavor assumptions as in Section \ref{sec:mixing}, the coupling matrices are given by
\begin{align}\label{eq:CoupMat}
\kappa_q&\approx
\left(\begin{array}{ccc:cc}
c_d^2\,(s_{q_1}^2 - c_{q_1}^2\,g_3^2/g_4^2)             & c_d^2\,|V_{td}/V_{ts}|\, (s_{q_1}^2 - s_{q_2}^2)    & 0                 &    -  c_{d} \,  c_{q_1} s_{q_1}&  |V_{td}/V_{ts}|\,c_{d} \,c_{q_{2}} s_{q_{2}}   \\[2pt]
c_d^2\,|V_{td}/V_{ts}|\, (s_{q_1}^2 - s_{q_2}^2)          & c_d^2\,(s_{q_2}^2 - c_{q_2}^2\,g_3^2/g_4^2)       &  0                     &     -|V_{td}/V_{ts}| \,  c_{d} \, c_{q_{1}} s_{q_{1}}                       & - c_d \, c_{q_2}\,s_{q_2}\\[2pt]
0                                                                                  & 0                                                               & 1                 &   0                           &  0 \\[2pt] \hdashline 
- c_{d} \,  c_{q_1}s_{q_1}                                                          &  -|V_{td}/V_{ts}| \,  c_{d} \, c_{q_{1}} s_{q_{1}}                                                              & 0                &    c_{q_1}^2             &  0 \\[2pt]
|V_{td}/V_{ts}| \,  c_{d} \, c_{q_{2}} s_{q_{2}}                                                                                 & -  c_{d} \, c_{q_2}\,s_{q_2}                                      & 0                &     0                          &  c_{q_{2}}^2\\
\end{array}\right)\,, \\[5pt]
\xi_q&\approx
\left(\begin{array}{ccc:cc}
c_d^2\,s_{q_1}^2                                                         & c_d^2\,|V_{td}/V_{ts}|\, (s_{q_1}^2 - s_{q_2}^2)    & 0                 &    - c_{d} \, c_{q_1} s_{q_1} &   |V_{td}/V_{ts}| \,  c_{d} \,c_{q_{2}} s_{q_{2}}   \\[2pt]
c_d^2\,|V_{td}/V_{ts}|\, (s_{q_1}^2 - s_{q_2}^2)    & c_d^2\,s_{q_2}^2                                                 &  0                 &     -|V_{td}/V_{ts}|  \, c_{d} \, c_{q_{1}} s_{q_{1}}                              & - c_{d}\,c_{q_2}\,s_{q_2}\\[2pt]
0                                                                       & 0                                                               & 1                 &   0                           &  0 \\[2pt] \hdashline 
- c_{d}\, c_{q_1}s_{q_1}                                               &  -|V_{td}/V_{ts}|  \,  c_{d} \, c_{q_{1}} s_{q_{1}}                                                      & 0                 &    c_{q_1}^2             &  0 \\[2pt]  
 |V_{td}/V_{ts}| \,  c_{d} \, c_{q_{2}} s_{q_{2}}                                                                     & - c_{d} \, c_{q_2}\,s_{q_2}                                      & 0                &     0                          &  c_{q_{2}}^2\\
\end{array}\right)\,,\\[5pt]
\xi_\ell&\approx
\left(\begin{array}{ccc:cc}
0              & 0                                    & 0             &    0      & 0                                     \\[2pt]
0              & s_{\ell_2}^2                    & -s_\tau    &    0       & -c_{\ell_2}\,s_{\ell_2}      \\[2pt]
0              & -s_\tau                            & 1             &    0       &   -s_{\tau}\,c_{\ell_2}\,s_{\ell_2}                                   \\[2pt] \hdashline 
0              & 0                                     & 0             &     1      &  0                                    \\[2pt]  
0              & -c_{\ell_2}\,s_{\ell_2}      & -s_{\tau}\,c_{\ell_2}\,s_{\ell_2}             &     0      &  c_{\ell_{2}}^2\\
\end{array}\right)\,,\\[5pt]
\kappa_u&\approx\kappa_d\approx\xi_u\approx\xi_d\approx\xi_e\approx\mathbb{1}_{3\times3}\,,\qquad\qquad \kappa_Q\approx\xi_Q\approx\xi_L\approx\mathbb{1}_{2\times2}\,,
\end{align}
where we neglected terms of $\mathcal{O}(g_1^2/g_4^2)$ and $\mathcal{O}(s_{q_{1,2}}\,g_3^2/g_4^2)$.  Note that the small breaking of $U(2)_q$ mentioned in Section~\ref{sec:mixing} has to do with the fact that $s_{q_1}\neq s_{q_2}$. From~\eqref{qlmix}, we can see that the difference between the two angles is sub-leading and therefore small enough to pass the stringent constraints from $D-\bar D$ mixing, see Section~\ref{sec:constraints}. We remind the reader that these interactions are given in the flavor basis for the $SU(2)_L$-doublets defined in~\eqref{eq:DownBasis}.

\bibliographystyle{JHEP}

{\footnotesize
\bibliography{paper}
}

\end{document}